\documentclass[manuscript,screen]{acmart}

\AtBeginDocument{%
  \providecommand\BibTeX{{%
    \normalfont B\kern-0.5em{\scshape i\kern-0.25em b}\kern-0.8em\TeX}}}

\setcopyright{acmcopyright}
\copyrightyear{2018}
\acmYear{2018}
\acmDOI{10.1145/1122445.1122456}

\acmConference[Woodstock '18]{Woodstock '18: ACM Symposium on Neural
  Gaze Detection}{June 03--05, 2018}{Woodstock, NY}
\acmBooktitle{Woodstock '18: ACM Symposium on Neural Gaze Detection,
  June 03--05, 2018, Woodstock, NY}
\acmPrice{15.00}
\acmISBN{978-1-4503-XXXX-X/18/06}

\usepackage{multirow}
\usepackage[ruled,vlined,linesnumbered]{algorithm2e}
\usepackage{xspace}
\usepackage{subcaption}
\usepackage{makecell}
\usepackage{multicol}
\usepackage{lipsum}
\usepackage{amsmath}
\usepackage{enumitem}

\setlist{leftmargin=5mm}

\theoremstyle{definition}

\definecolor{green}{rgb}{0.0, 0.27, 0.13}
\definecolor{applegreen}{rgb}{0.55, 0.71, 0.0}

\newcommand{\name}{\textsc{Hybiscus}\xspace}
\newcommand{\dfj}{\textsc{Defects4J}\xspace}
\newcommand{\sir}{\textsc{SIR}\xspace}

\newcommand{\hdist}{\textit{hdist}\xspace}

\newcommand{\exprrepo}{\url{https://github.com/anonytomatous/Hybiscus}}
\newcommand{\resultlink}{\url{https://github.com/anonytomatous/Hybiscus/blob/flattened/results.md}}

\newcommand{\javarepo}{\url{https://github.com/anonytomatous/docker-D4J-multifault}}
\newcommand{\crepo}{\url{https://github.com/anonytomatous/docker-SIR-multifault}}

\begin{document}

\title{Improving Test Distance for Failure Clustering with Hypergraph Modelling}

\author{Gabin An}
\affiliation{%
  \institution{KAIST}
  \city{Daejeon}
  \country{Republic of Korea}
}
\email{agb94@kaist.ac.kr}

\author{Juyeon Yoon}
\affiliation{%
  \institution{KAIST}
  \city{Daejeon}
  \country{Republic of Korea}
}
\email{juyeon.yoon@kaist.ac.kr}

\author{Joyce Jiyoung Whang}
\affiliation{%
  \institution{KAIST}
  \city{Daejeon}
  \country{Republic of Korea}
}
\email{jjwhang@kaist.ac.kr}

\author{Shin Yoo}
\affiliation{%
  \institution{KAIST}
  \city{Daejeon}
  \country{Republic of Korea}
}
\email{shin.yoo@kaist.ac.kr}

\begin{abstract}
Automated debugging techniques, such as Fault Localisation (FL) or Automated Program Repair (APR), are typically designed under the Single Fault Assumption (SFA). However, in practice, an unknown number of faults can independently cause multiple test case failures, making it difficult to allocate resources for debugging and to use automated debugging techniques. Clustering algorithms have been applied to group the test failures according to their root causes, but their accuracy can often be lacking due to the inherent limits in the distance metrics for test cases. We introduce a new test distance metric based on hypergraphs and evaluate their accuracy using multi-fault benchmarks that we have built on top of \dfj and \sir. Results show that our technique, \name, can automatically achieve perfect clustering (i.e., the same number of clusters as the ground truth number of root causes, with all failing tests with the same root cause grouped together) for 418 out of 605 test runs with multiple test failures. Better failure clustering also allows us to separate different root causes and apply FL techniques under SFA, resulting in saving up to 82\% of the total wasted effort when compared to the state-of-the-art technique for multiple fault localisation. 

\end{abstract}

\maketitle

\section{Introduction}
\label{sec:introduction}

As software systems grow in size and complexity, the cost of debugging has significantly 
increased. Many automated techniques have been proposed and studied to reduce the burden of debugging.
Fault Localisation (FL) aims to automatically identify the location of the root cause of the observed test failure~\cite{Wong:2016aa}, using various information such as program spectrum~\cite{jones2005empirical, naish2011model, wong2013dstar}, mutation analysis~\cite{moon2014ask, hong2017museum, papadakis2012using, papadakis2015metallaxis}, and textual similarity between bug reports and source code~\cite{zhou2012should, saha2013improving, lukins2008source}. Automated Program Repair (APR) uses results of FL to identify the location of the fault, and seeks to generate patches, either by finding ingredients of the patch from existing code~\cite{Yuan2018arja,Wen2018dk}, or by synthesising the patch based on the observed violation of oracles~\cite{Le2017vo,Mechtaev2016rw}. 

Most FL and APR techniques that have made significant advances share a common basis, 
which is the Single Fault Assumption (SFA): they assume that there exists a single fault in the
System Under Test (SUT) that is responsible for all observed test failures. SFA allows us to 
precisely measure the effectiveness of FL and APR techniques, which in turn enables the design of more 
advanced automated debugging techniques.
Fault benchmarks such as \dfj~\cite{just2014defects4j} 
contain significant amounts of effort to capture and reproduce real-world faults in isolation, so 
that automated techniques can be studied and developed under SFA.

In practice, however, SFA does not 
always hold. It is entirely possible that a set of changes made to SUT contains multiple faults, each
being the root cause of different test cases. Multiple faults that occur simultaneously present 
challenges not only to automated debugging techniques developed under SFA~\cite{digiuseppe2015fault, xue2013significant, digiuseppe2011influence},
but also to human engineers whose very first task is to understand how many faults there are to debug.

Given multiple test failures, how can we decide the number of different root causes, as well as the 
mapping between the causes and the observed failures? With any SUT of realistic complexity, the 
number of failing test cases may not directly indicate the number of different root causes, as the 
dependency structures in SUT can force a single fault to affect the outcomes of multiple test
cases. Clustering of test cases has been proposed as a solution to group failing test cases~\cite{jones2007debugging, golagha2017reducing, gao2017mseer,golagha2019failure}, but the 
accuracy of clustering, in terms of both the number of clusters (i.e., the number of root causes) and the cluster 
membership (i.e., the mapping between root causes and test failures) can be lacking. Inaccurate clustering would 
introduce additional challenges to the debugging process, due to inefficient resource management based on incorrect 
estimation of root causes and the sub-optimal performance of automated debugging techniques.

We propose \name,
a \textbf{hy}pergraph \textbf{b}ased fa\textbf{i}lure repre\textbf{s}entation and \textbf{c}l\textbf{us}tering technique
that can accurately predict both the number of 
root causes and the mapping between root causes and test failures. \name introduces a hypergraph
representation of test coverage, from which it also derives a novel test distance metric.
In addition to \name, we also introduce multiple-fault variants of widely studied fault benchmarks,
\dfj and \sir: both were constructed by systematically merging real (\dfj) and seeded (\sir)
faults in the original benchmarks. Our empirical evaluation shows that, when used with Agglomerative
Hierarchical Clustering (AHC) and a distance-based estimation of cluster numbers,
\name can significantly outperform other failure clustering methods. Once clustered, \name can apply
total wasted effort when compared to the state-of-the-art multi-fault FL technique, MSeer~\cite{gao2017mseer}.

The main contribution of this paper includes the following.

\begin{itemize}
\item We propose a novel test distance metric, \hdist, based on hypergraph modelling.
Our hypergraph based distance metric can measure the dissimilarity 
between two test cases while reflecting their their higher-order relationship 
with the remainder of the test suite. 

\item We introduce multiple fault variants of the \dfj and \sir
benchmarks to evaluate our novel test distance metric in the context of 
failure clustering. Each of the faulty versions in our variants includes up to 
seven distinct faults. Our datasets for Java\footnote{\javarepo} and C\footnote{\crepo} subjects are publicly available and can be used for future research
on automated debugging of multiple faults. To the best of our knowledge,
there has been no failure clustering work validated on both Java and C 
subjects.

\item The empirical evaluation using the multiple fault versions of \dfj and 
\sir shows that our novel hypergraph-based test distance metric can more 
accurately measure the distance between test cases when compared to other 
vector-, set-, and ranking-based metrics in the application on failure 
clustering.

\item We introduce \name\footnote{\exprrepo}, a failure clustering technique
that uses the hypergraph-based test distance. Using heuristic estimation of 
cluster numbers, \name can perfectly cluster 69\% of the studied multiple 
fault subjects, i.e., with the correct number of root causes and correct 
groupings of failing test cases, without any human intervention. Using the 
results of clustering, \name can also localise each of the multiple faults 
saving up to 82\% of the total wasted effort when compared to the 
state-of-the-art multiple fault localisation technique, MSeer.

\end{itemize}

The paper is structured as follows. Section~\ref{sec:background} formally introduces the problem of
failure clustering, with a motivating example. Section~\ref{sec:hybiscus} presents the hypergraph
representation of test coverage, as well as the distance metric between test coverage and our
clustering formulation. Section~\ref{sec:experimental_setup} describes how multiple fault variants
of \dfj and \sir were constructed and presents our experimental setup. Section~\ref{sec:results}
presents the results of our empirical evaluation, and Section~\ref{sec:threats} discusses threats to validity. Section~\ref{sec:related_work} presents related work, and Section~\ref{sec:conclusion} concludes.

\section{Failure Clustering}
\label{sec:background}

Let us consider a program that consists of $M$ components, $C = \{c_1, c_2, \ldots, c_M\}$, and a test suite 
with $N$ test cases,  $T = \{t_1, t_2, \ldots, t_N\}$. After executing $T$, let $T_P$ and $T_F$ be the set of passing and failing test cases, respectively (note that the test 
coverage can be represented by a $N \times M$ matrix $A$, whose entry $A_{i,j}$ is 1 if $t_i$ executes 
$c_j$ and 0 otherwise). When there are multiple failing test cases, i.e., $|T_F| > 1$, a developer
should separate different failures according to their root causes in order to debug them one by one.

The problem of \textit{failure clustering} is to 
assign cluster membership to failing test cases in $T_F$ so that, in the 
resulting clusters, $P = \{T_1, \ldots, T_k\}$, failures due to the same 
root cause are grouped together.\footnote{We perform non-overlapping clustering,
based on the failure-to-single-fault assumption. That is, we assume that each failing test case has
one root cause. For discussion of our future work on overlapping failure clustering,
see Section~\ref{sec:threats}.}
For clustering problems for which ground-truth cluster assignment $P_{true}$ is known,
there are two desirable properties of a good cluster assignment $P$~\cite{rosenberg2007v}:

\begin{itemize}
\item \textbf{Homogeneity}: Every member of a cluster in $P$ is assigned to the same cluster in $P_{true}$. 
\item \textbf{Completeness}: Every member of a cluster in $P_{true}$ is assigned to the same cluster in $P$.
\end{itemize}

For the failure clustering problem, a perfect cluster assignment is the one that assigns each failing test in $T_F$ to groups that share the same root cause: each cluster in $P$ has its own root cause, which is different from those 
of all other clusters. In this context, we can rephrase homogeneity and completeness as follows:

\begin{itemize}
\item All failing test cases in a cluster share the same root cause (homogeneity).
\item All failing test cases sharing the same root cause belong to the same cluster (completeness).
\end{itemize}

If a failure clustering is not homogeneous, anyone using one of the clusters to understand one of the root 
causes will be misled, as the non-homogeneous test failures will add noise to the process. If a failure
clustering is not complete, there will be redundant clusters, which subsequently will increase the inspection 
workload for the developer. Finding both the correct assignment of failing test cases to root causes
and the correct number of root causes is important for effective and efficient debugging of multiple faults.

\subsection{A Motivating Example}
\label{sec:background:moti}

Let us consider a System Under Test (SUT) with six components, the set of which is denoted by 
$C = \{c_1, \ldots, c_6\}$. Let $T$ be a test suite with five test cases, $T = \{t_1, \ldots, t_5\}$.
Suppose there are two faulty components, $c_3$ and $c_4$: test case $t_3$ fails due to the execution of $c_3$, 
while $t_4$ and $t_5$ fail due to $c_4$. The full execution traces are shown in 
Table~\ref{tab:trace_ex}. Our goal is to cluster the set of failing test cases, $T_F = \{t_3, t_4, t_5\}$, 
into $P = \{\{t_3\}, \{t_4, t_5\}\}$.

\begin{table}[t]
\renewcommand{\arraystretch}{1.0}
\caption{Example of test execution traces}
\label{tab:trace_ex}
\centering
\scalebox{0.85}{

\begin{tabular}{c|c|c|c|c|c}
\toprule
TC&$t_1$&$t_2$&$t_3$&$t_4$&$t_5$\\\midrule
Trace&$\{c_1, c_2, c_5, c_6\}$&$\{c_1, c_2, c_5, c_6\}$&$\{c_2, \underline{\mathbf{c_3}}, c_5\}$&$\{c_1, \underline{\mathbf{c_4}}\}$&$\{c_2, \underline{\mathbf{c_4}}, c_5\}$\\\midrule
Result&{\color{green}Pass}&{\color{green}Pass}&{\color{red}Fail}&{\color{red}Fail}&{\color{red}Fail}\\
\bottomrule
\end{tabular}
}
\end{table}

We can expect that failing test cases with similar behaviour are more likely to share the same root cause. Since 
execution traces reflect test case behaviour, we can cluster failing test cases using a distance metric 
defined over the execution traces.
In the next section, we discuss distance metrics and the test execution trace representations
used in the literature to perform failure clustering.

\subsection{Distance Metrics for Failure Clustering}
\label{sec:distance_metrics}

Various distance metrics have been proposed for failure clustering to capture
the proximity between observed test failures~\cite{liu2008systematic}. The 
representation of test execution guides the choice of a distance metric, which in turn affects the performance
of failure clustering~\cite{zakari2020multiple}. The most widely studied representations of test execution traces are numerical vectors, sets, and fault localisation ranks.

\subsubsection{Numerical Vectors}

A test execution trace can be represented as a $M$-dimensional vector. When the vector is binary, this 
representation becomes the coverage vector, 1 meaning the corresponding component being covered, and
0 otherwise. For example, the test case $t_3$ in Section~\ref{sec:background:moti} can be represented
by $(0, 1, 1, 0, 1, 0)$. For binary vector representation, Hamming, Cosine, or Euclidean distance
can be used to measure distances between test cases~\cite{yoo2009clustering, huang2013empirical, wang2014weighted, golagha2017reducing}.
Beyond binary representation, each vector element could also contain the number of times the
corresponding component has been executed. However, to the best of our knowledge,
most existing failure clustering literature focuses on binary representation.

While intuitive, previous literature~\cite{gao2017mseer, liu2008systematic} argues that this
representation is inappropriate for failure clustering, because test cases that fail due to the
same root cause may still produce considerably different execution traces.
For example, $t_4$ and $t_5$ in Section~\ref{sec:background:moti} only share one program component,
$c_4$, which is the root cause.
Using Hamming and Euclidean distance metrics, the distances between the binary vector
representations of $t_3$, $t_4$ and $t_5$ are as follows:

\begin{itemize}
\item $Hamming(t_3, t_4) = 0.83$ and $Euclidean(t_3, t_4) = 2.24$
\item $Hamming(t_4, t_5) = 0.50$ and $Euclidean(t_4, t_5) = 1.73$
\item $Hamming(t_3, t_5) = \textbf{0.33}$ and $Euclidean(t_3, t_5) = \textbf{1.41}$
\end{itemize}

According to these results, $t_3$ and $t_5$ are the closest among all pairs of failing test cases.
Consequently, the pair of $t_3$ and $t_5$ is more likely to be assigned to the same cluster than
that of $t_4$ and $t_5$ by any clustering algorithm that uses these distance metrics. However,
this is not aligned with the ground-truth clustering: $\{\{t_3\}, \{t_4, t_5\}\}$.

\subsubsection{Sets}

An execution trace can also be represented as a set of all program components it covers. There are many set 
similarity metrics that are widely used, such as Jaccard or S{\o}rensen-Dice coefficient.
However, the set representation can be vulnerable to variance in execution traces
of failing tests, such like the vector representation.
Consider the set-based distances between failing test cases in the motivating example:

\begin{itemize}
\item $Jaccard(t_3, t_4) = 1.00$ and $Dice(t_3, t_4) = 1.00$
\item $Jaccard(t_4, t_5) = 0.75$ and $Dice(t_4, t_5) = 0.66$
\item $Jaccard(t_3, t_5) = \textbf{0.50}$ and $Dice(t_3, t_5) = \textbf{0.33}$
\end{itemize}

As in the case of the vector representation, the set-based distance metrics pronounce that $t_3$ and $t_5$ is the
closest pair failing test cases. The common weakness of both vector and set representation is that it is 
difficult to capture the due-to relationship between test failures and their root causes~\cite{gao2017mseer}.
This is because both representations put equal importance to all program components. For example, when calculating
the distance between $t_4$ and $t_5$, we should put more weight on the program component $c_4$ since it is 
executed by only failing test cases, thus more suspicious. However, such globally available information is not 
reflected in these representations.

\subsubsection{Ranking} 
\label{sec:background:distance:ranking}

A \textit{failing} test case can be represented as a suspiciousness ranking list. Unlike vectors and
sets that can represent any execution traces, the ranking-based representation~\cite{liu2008systematic, jones2007debugging, gao2017mseer} 
was specifically designed to cluster failing test cases, while considering passing test cases as well.
A failing test case is represented as a ranking of $M$ program components in descending order of
their suspiciousness scores. The suspiciousness scores are, in turn, computed by applying an FL technique to the subset
of all passing test cases plus the failing test case under consideration. Once all failing test cases receive 
their ranking, distances between them can be computed using metrics such as (Revised) 
Kendall-Tau (RKT) distance~\cite{gao2017mseer}. Compared to vectors and sets, the ranking-based representation 
focuses on the failures through the use of FL techniques, which also utilise passing test cases in their analysis. 

Let us compute the RKT distance for our motivating example. Using Crosstab~\cite{wong2008crosstab} as the FL technique, following Gao and Wong~\cite{gao2017mseer}, produces the following distances:

\begin{itemize}
    \item RKT(t3, t4) = 19.72, RKT(t4, t5) = \textbf{2.92}, RKT(t3, t5) = 16.80
\end{itemize}

In contrast to vectors and sets, RKT can correctly predict that $t_4$ and $t_5$ are the closest pair of failing 
test cases. However, ranking-based representation is not without any weaknesses. The choice of fault localisation
technique, as well as the choice of tie-breaking schemes, can affect the performance of failure clustering. Both 
Kendall-Tau and revised Kendall-Tau distance are also computationally expensive with a complexity of 
$O(M^2|T_F|^2)$, as they compare the relative differences in ranks for every program element pair to calculate 
the distance between two failing test cases (we report the significant computational cost analysis of RKT in 
Section~\ref{sec:result:RQ3}).

\section{\name: Failure Clustering using Hypergraph-based Test Distance}
\label{sec:hybiscus}

We model the test coverage of a faulty program as a hypergraph.
Then, the original failure clustering problem is converted
into a hypergraph clustering problem, and the distance between hypergraph
vertices acts as a proxy of the distance between test cases.
In the next section, we formally define the basic notation of hypergraph.

\subsection{Hypergraphs}
\label{sec:failure_clustering:hypergraph}
A hypergraph is a graph whose edges can join any number of vertices, not only two vertices.
These edges are called \textit{hyperedges}, and a hyperedge is appropriate for modelling higher-order
relationships among objects~\cite{zhou2006learning}.

Formally, a hypergraph $G = (V, E, w)$ is a triplet of a set of
vertices $V = \{v_1, v_2, ..., v_N\}$,
a set of hyperedges $E = \{e_1, e_2, ..., e_M\}$
, and a function $w \in E \rightarrow \mathbb{R}_{0+}$ that maps a hyperedge $e \in E$ to its non-negative weight, $w(e)$, following the notation of Zhou et al.~\cite{zhou2006learning}.
A hyperedge $e$ is represented as a subset of
vertices that $e$ connects, and the union of all hyperedges are equal
to the set of all vertices,
i.e., $\forall e \in E, e \subseteq V$ and $\bigcup_{i=1}^{M} e_i = V$.
The degree of a vertex $v$ is defined by $deg(v) = \sum_{\{e \in E|v \in e\}}w(e)$, and the degree of a hyperedge $e$ is defined by $deg(e) = |e|$.

An $N \times M$ incidence matrix $H$ represents the vertex-hyperedge
relationships in $G$. $H_{ij}$ is $1$ if the vertex $v_i$ is included in the
hyperedge $e_j$ and 0 otherwise.
Also, let $W \in \mathbb{R}_{+}^{M \times M}$ and $D_e \in \mathbb{R}_{+}^{M \times M}$
denote the weight and degree diagnonal matrices such that
$W_{j,j} = w(e_j)$ and $[D_e]_{j,j} = deg(e_j)$.

\subsection{Hypergraph Modelling of Test Coverage}
\label{sec:failure_clustering:modelling}

\begin{figure}[t]
\centering
\centerline{\includegraphics[width=0.5\textwidth]{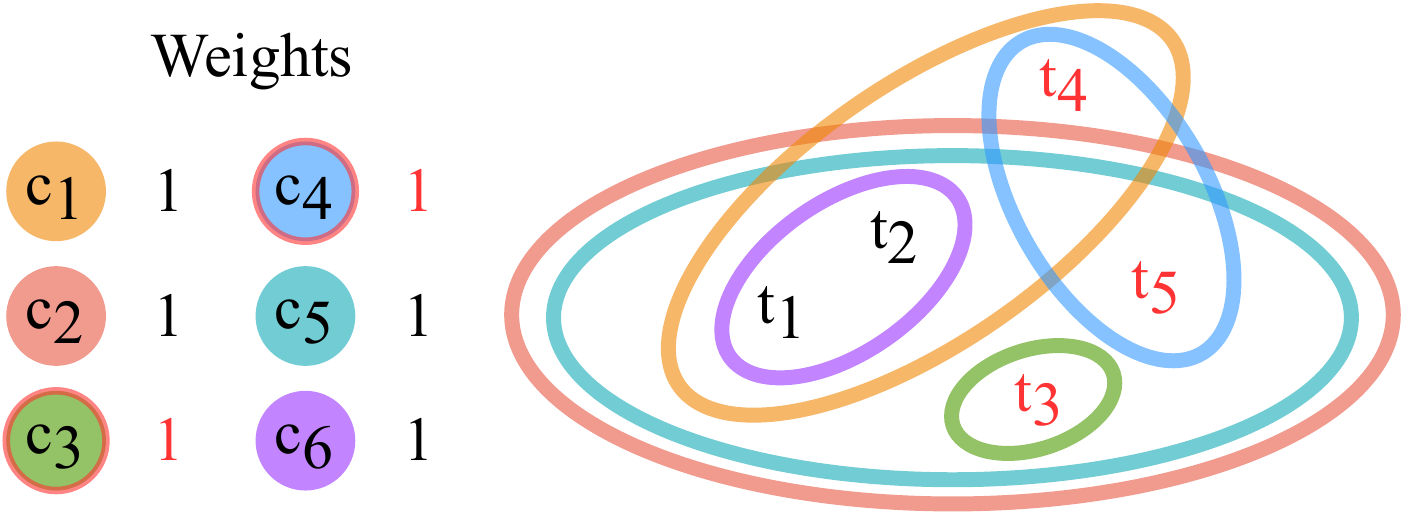}}
\caption{Hypergraph modelling of test coverage in Table~\ref{tab:trace_ex}. Faulty components and failing test cases are marked in red.}
\label{fig:hypergraph}
\end{figure}

Suppose that we convert the execution traces in Table~\ref{tab:trace_ex} to a
hypergraph whose vertices and hyperedges correspond to test cases and program
components, respectively.
As shown in Figure~\ref{fig:hypergraph}, let each program component (hyperedge)
connect all the test cases (vertices) which executed the program component.
For example, $c_1 = \{t_1, t_2, t_4\}$ and $c_6 = \{t_1, t_2\}$.
The incidence matrix of the hypergraph in Figure~\ref{fig:hypergraph} is then identical to the coverage matrix of the program under modelling:
\begin{equation}
\label{eq:incidence}
H=\left[\begin{array}{cccccc}
        1 & 1 & 0 & 0 & 1 & 1\\
        1 & 1 & 0 & 0 & 1 & 1\\
        0 & 1 & 1 & 0 & 1 & 0\\
        1 & 0 & 0 & 1 & 0 & 0 \\
        0 & 1 & 0 & 1 & 1 & 0 \\
\end{array}\right]
\end{equation}

Thus, the hypergraph can model the test coverage without the loss of information.

Formally, given a set of program components $C$, a test suite $T$, and a coverage matrix $A$,
we construct a hypergraph $G = (V, E, w)$ such that $V = T$, $E = \{\{t_i \in T|A_{ij} = 1\}|c_j \in C\}$, $\forall e \in E. w(e)=1$.
Then, the incidence matrix $H$ of $G$ satisfies $H = A$.\footnote{We suppose 
that $T$ do not include test cases not covering any program component.}

\subsection{Defining Distance using a Hypergraph}
\label{sec:failure_clustering:hyper_dist}

There are several ways to measure the proximity between hypergraph vertices.
In this work, referring to the recent article on hypergraph clustering~\cite{whang2020mega},
we first define the linkage between two vertices $v, v' \in V$ by:
\begin{equation}
l(v,v') = \sum_{e \in E_{v} \cap E_{v'}}\frac{w(e)}{deg(e)}
\label{eq:link}
\end{equation}
where $E_{v} = \{e\in E|v \in e\}$.
As a hyperedge in $E_{v} \cap E_{v'}$ connects fewer vertices and has the greater weight value,
the linkage value will be higher; the linkage value encodes how \textit{tightly} the two vertices are connected.

Since $E_{v} \cap E{v'} \subseteq E_{v}$,
\begin{equation*}
l(v,v') = \sum_{e \in E_{v} \cap E_{v'}}\frac{w(e)}{deg(e)} \leq \sum_{e \in E_{v}}\frac{w(e)}{deg(e)} = l(v, v)
\label{eq:bound}
\end{equation*}
Let us define $assoc(v) := l(v,v)$. Similarly, $l(v,v') \leq assoc(v')$.
Therefore, we can normalise the linkage value in Eq.~\ref{eq:link} as follows:
\begin{equation}
\hat{l}(v, v') = \frac{1}{2}\left(\frac{l(v,v')}{assoc(v)} + \frac{l(v,v')}{assoc(v')}\right)
\label{eq:nlink}
\end{equation}
Then, $0 \leq \hat{l}(v, v') \leq 1$. We can simply prove that $\hat{l}(v,v') = 1$ when $E_{v} = E_{v'}$, and $\hat{l}(v,v') = 0$ when $E_{v} \cap E_{v'} = \emptyset$.
This normalisation aims to measure the \textit{relative importance} of the linkage between the
two vertices, $v$ and $v'$, compared to the total association of the vertices.
There is another normalisation method dividing the linkage value by $deg(v)$~\cite{whang2020mega}, not $assoc(v)$.
However, we choose the denominator as $assoc(v)$ since it is a tighter upper bound on $l(v, v')$ than $deg(v)$.
Also, through our initial experiments, we have found that using $assoc(v)$ is more effective in failure clustering
that using $deg(v)$ as a normalisation denominator.

The normalised linkage can be calculated in matrix form:
\begin{equation}
\hat{L} = \frac{1}{2}\left((L \odot I)^{-1}L + L(L \odot I)^{-1}\right)
\label{eq:nlink_matrix}
\end{equation}
where $L = HWD_{e}^{-1}H^{T}$ is an unnormalised linkage matrix, and $\odot$ means the element-wise product. The matrix elements $L_{i,j}$ and
$\hat{L}_{i,j}$ correspond to $l(v_i, v_j)$ and $\hat{l}(v_i, v_j)$, respectively.

Let us recall the motivating example in Section~\ref{sec:background:moti},
which is modelled as a hypergraph in Figure~\ref{fig:hypergraph}.
The normalised linkage values between the all pairs of vertices corresponding to
failing tests, $T_F = \{t_3, t_4, t_5\}$, are as follows:
$$
\hat{l}(t_3, t_4) = 0, \hat{l}(t_4, t_5) = \textbf{0.55}, \hat{l}(t_3, t_5) = 0.42
$$

In contrast to the vector- or set-based distance metrics in Section~\ref{sec:distance_metrics},
the linkage values tell us that $t_4$ and $t_5$ has a stronger relationship than
$t_3$ and $t_5$. This is because it incorporates the hyperedge degrees, which
reflect the global coverage information.
The more common a program component is, the more loosely the corresponding hyperedge is
considered to connect the vertices.

Finally, we define the distance between the vertices based on the normalised linkage $\hat{l}$ (Eq.~\ref{eq:link}) by:
\begin{equation}
\textit{hdist}(v, v') = 1 - \hat{l}(v, v')
\label{eq:hdist}
\end{equation}

As the linkage between vertices is stronger, the distance is smaller.
For example, $t_3$ and $t_4$ (in Figure~\ref{fig:hypergraph}) do not share
any hyperedges, so $\textit{hdist}(t_3, t_4) = 1$, which is the maximum distance value.
For all hypergraph vertices, the distance from itself is zero.

\subsection{Subgraph Extraction using Failing Tests}

Our original purpose is to cluster the set of failing test cases, i.e., 
to cluster a \textbf{subset} of vertices corresponding to failing test cases, $T_F$, and 
not the entire vertices $V=T$. 
Consequently, we eliminate the vertices of passing tests from the hypergraph
while still preserving their information; this can reduce the computational cost of Eq.~\ref{eq:nlink_matrix}.
To do so, we ensure that this elimination does not affect the distance
between failing test cases by readjusting the weights of hyperedges.

Formally, given an original hypergraph $G=(V, E, w)$ modelling the test coverage,
we reconstruct a hypergraph by including only vertices that correspond to failures and their adjacent hyperedges:
\begin{equation}
\label{eq:restriction}
G_{T_F} = (T_F, \{e \cap T_F|e \in E \wedge e \cap T_F \neq \emptyset \}, w')
\end{equation}
We call this subhypergraph as a restriction of $G$ to $T_F$.

In Section~\ref{sec:failure_clustering:hyper_dist},
\textit{hdist} is defined in terms of $\hat{l}$, and $\hat{l}$ is defined in terms of $l$.
Therefore, if we preserve $l$ between every pair of failing vertices, the distance does not change.
Recall that the linkage $l(v,v')$ is the sum of the ratio between the weight and the degree
over all hyperedges connecting $v$ and $v'$. 
Since the set of hyperedges among failing test vertices are the same in $G_{T_F}$,
the only value needed to be preserved in $G_{T_F}$ is the ratio.

While the elimination of vertices may lead to the decrease in the degree of some hyperedge $e$,
let $deg(e)$ be the degree of $e$ in $G$ and $deg'(e)$ be the new degree of $e$ in $G_{T_F}$.
Then, to preserve the ratio value, the new hyperedge weight $w'(e)$ for all hyperedges $e$ in $G_{T_F}$
should satisfy the following equation:%
\begin{equation}
\frac{w'(e)}{deg'(e)}=\frac{w(e)}{deg(e)}
\label{eq:weight_condition}
\end{equation}

Since the original weights of all hyperedges are initially set to $1$ (in Section~\ref{sec:failure_clustering:modelling}), Eq.~\ref{eq:weight_condition} is equivalent to
\begin{equation}
w'(e) = \frac{deg'(e)}{deg(e)} (\because \forall e. w(e) = 1)
\label{eq:f}
\end{equation}

Mathematically, $w'(e_j)$ is equivalent to $\frac{|\{t_i \in T_F|A_{ij} = 1\}|}{|\{t_i \in T|A_{ij} = 1\}|}$,
which is the ratio of failing test cases among all test cases covering the corresponding program component $c_j$.

\begin{figure}[t]
\centering
\centerline{\includegraphics[width=0.4\textwidth]{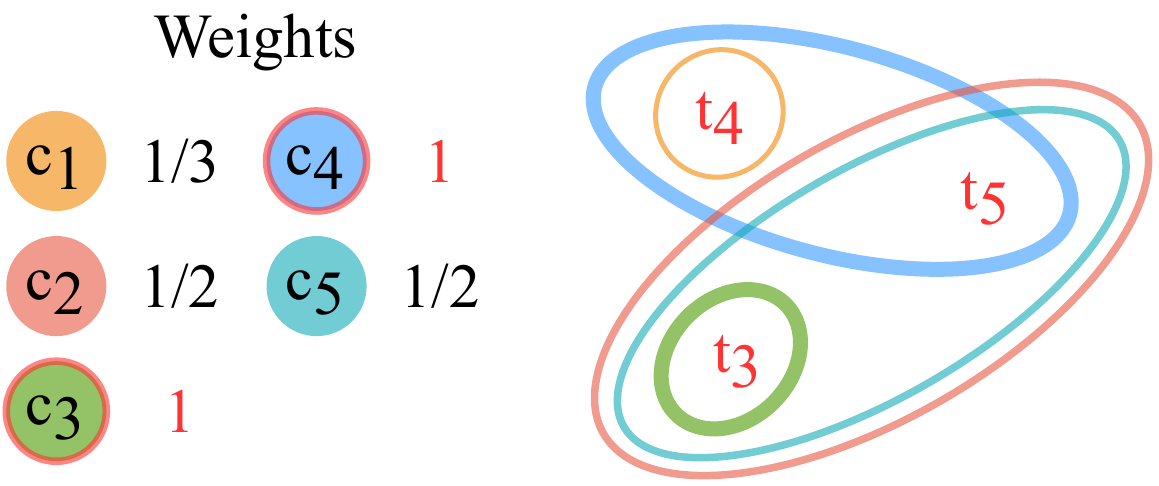}}
\caption{Restriction of the hypergraph in Figure~\ref{fig:hypergraph} to vertices corresponding to failing tests.
The thickness of each hyperedge is proportional to its readjusted weight.}
\label{fig:restricted_hypergraph}
\end{figure}

Figure~\ref{fig:restricted_hypergraph} shows the restriction of the hypergraph in 
Figure~\ref{fig:hypergraph} with the readjusted weights of the hyperedges,
$(\frac{1}{3}, \frac{1}{2}, 1, 1, \frac{1}{2})$.
The incidence matrix of this reduced hypergraph is a submatrix of $H$
(in Eq.~\ref{eq:incidence}) formed by the rows and the columns
that correspond to the remaining vertices and hyperedges, respectively:
$$
H_{3:5,1:5}=\left[\begin{array}{ccccc}
        0 & 1 & 1 & 0 & 1 \\
        1 & 0 & 0 & 1 & 0 \\
        0 & 1 & 0 & 1 & 1 \\
\end{array}\right]
$$

Finally, the pairwise distance matrices, i.e. $1 - \hat{L}$, before
and after the restriction are as follows:
$$
\left[\begin{array}{ccccc}
    0.00 & 0.00 & 0.53 & 0.60 & 0.45\\
    0.00 & 0.00 & 0.53 & 0.60 & 0.45\\
    0.53 & 0.53 & \textbf{0.00} & \textbf{1.00} & \textbf{0.58}\\
    0.60 & 0.60 & \textbf{1.00} & \textbf{0.00} & \textbf{0.45}\\
    0.45 & 0.45 & \textbf{0.58} & \textbf{0.45} & \textbf{0.00}\\
\end{array}\right] \rightarrow
\left[\begin{array}{ccc}
    \textbf{0.00} & \textbf{1.00} & \textbf{0.58}\\
    \textbf{1.00} & \textbf{0.00} & \textbf{0.45}\\
    \textbf{0.58} & \textbf{0.45} & \textbf{0.00}\\
\end{array}\right]
$$
The restriction does not alter the distance between failing test cases.

\subsection{Agglomerative Hierarchical Clustering}

We use a hierarchical clustering algorithm instead of partitional clustering
algorithms such as K-means~\cite{kanungo2002efficient}. When the number of 
faults is not known in advance, the freedom to derive any number of clusters 
from a single application of hierarchical clustering algorithm can be 
beneficial~\cite{golagha2019failure}. Users can examine the resulting 
dendrogram using their domain knowledge and find an appropriate stopping point.
Many existing failure clustering techniques also use hierarchical clustering due to the same reasons~\cite{jones2007debugging, golagha2017reducing, golagha2019failure}.

We follow the typical Agglomerative Hierarchical Clustering (AHC) process: it starts with each failing test case
being set as an individual cluster and merges the two \textit{closest} clusters at each iteration.
When the distance between two tests $t$ and $t'$ is given as $dist(t, t')$,
there are several ways to define the intercluster distance between two clusters $C$ and $C'$.
In this work, we compare the three linkage methods defined in Table~\ref{tab:intercluster_dist}
that aggregate the distance between all vertex pairs in $C$ and $C'$. Note that any pairwise test distance
function including \textit{hdist} can be plugged into $dist$.

\begin{table}[t]
\caption{Different options of the intercluster distance}
\label{tab:intercluster_dist}
\centering
\begin{tabular}{c|c}
\toprule
Name & Definition\\\midrule
Average (avg) & $D_{dist}(C, C') = \frac{1}{|C||C'|}\sum_{t \in C, t' \in C'}dist(t, t')$\\\hline
Single (min) & $D_{dist}(C, C') = \min_{t \in C, t' \in C'}dist(t, t')$\\\hline
Complete (max) & $D_{dist}(C, C') = \max_{t \in C, t' \in C'}dist(t, t')$\\
\bottomrule
\end{tabular}
\end{table}

Algorithm~\ref{algo:aggl} formally depicts the agglomerative clustering process.
In our approach, we set the input parameter $D_{dist}$ to $D_{hdist}$.
As mentioned earlier, the algorithm begins with $|T_F|$ clusters in which each failing test case is set to an
individual cluster (Line 2). Then, in each iteration, the nearest clusters are combined (Line 4-6).
This process continues until the number of clusters becomes $1$ (Line 3). Finally, it returns
the clustering results of test cases and the minimum distance results from every iteration (Line 9). 
Note that $P_k$ denotes the size $k$ partition of $T_F$, and $mdist_{k}$ is the minimum intercluster
distance when there are $k$ clusters.

\begin{algorithm}[t]
\small
\SetAlgoLined
\KwIn{Failing test cases $|T_F|$, Intercluster distance function $D_{dist}$}
\KwOut{Clustering results $P_1$, \dots, $P_N$, Mininum distance results $mdist_2$, \dots, $mdist_{N}$}
  $N \leftarrow |T_F|, k \leftarrow |T_F|$\\
  $P_k \leftarrow \{\{t\}|t \in T_F\}$\\
  \While{$k > 1$}{
    $C_1, C_2 \leftarrow argmin_{C \in P_k, C' \in P_k, C \neq C'} D_{dist}(C, C')$\\
    $mdist_{k} \leftarrow D_{dist}(C_1, C_2)$\\
    $P_{k-1} \leftarrow (P_{k} \setminus \{C_1, C_2\}) \cup \{C_1 \cup C_2\}$\\
    $k \leftarrow k - 1$
  }
  return $\{P_1, \dots, P_N\}$, $\{mdist_2, \dots, mdist_{N}\}$
 \caption{Agglomerative clustering of failing test cases}
 \label{algo:aggl}
\end{algorithm}

\begin{figure}[t]
\centering
\includegraphics[width=0.5\textwidth]{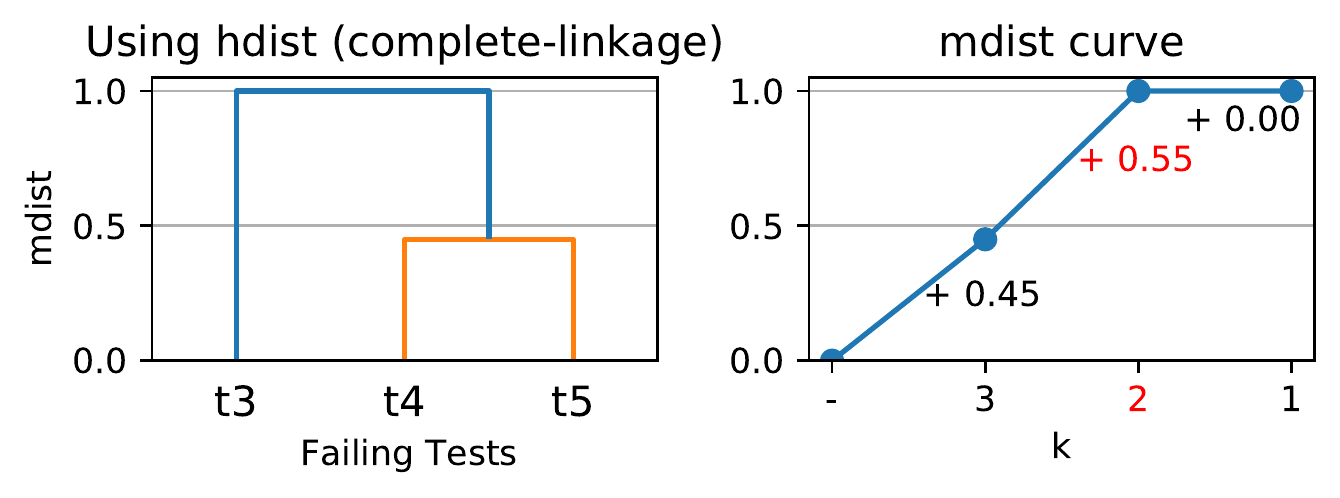}
\caption{Dendrogram (left) and $mdist$ curve (right) for the failure clustering results of Figure~\ref{fig:restricted_hypergraph} (\textit{hdist}-complete used), Since $mdist_k$ can be defined only when $k > 1$, $mdist_1$ is simply set to $1$, which is the upper bound of \textit{hdist}.}
\label{fig:dendrogram_example}
\end{figure}

For example, if we run this algorithm on the previous example (Figure~\ref{fig:restricted_hypergraph})
with the \textit{Complete} intercluster distance,
$P_3$ and $P_2$ will be $\{\{t_3\}, \{t_4\}, \{t_5\}\}$ and $\{\{t_3\}, \{t_4, t_5\}\}$, respectively,
while $mdist_3=0.45$ and $mdist_2=1.0$.
The clustering results can be represented as a dendrogram as shown in Figure~\ref{fig:dendrogram_example}.
The curve of $mdist$ according to the number of clusters $K$ can be deduced from the dendrogram.
If we cut this dendrogram
at the distance threshold $0.8$, the number of clusters is two, and
the failing test cases can be perfectly clustered.

Although users can manually decide the number of clusters from the dendrogram,
the clustering tool would be more useful if it can assist users by automatically suggesting
the proper number of clusters.
The simplest way might be setting the distance threshold to a fixed value
that works best empirically. On the other hand, in previous work~\cite{jones2007debugging, golagha2017reducing},
the stopping criterion is defined based on the fault localisation results.

In this work, we choose a stopping point using the \textit{elbow method} which is widely 
used to determine the number of clusters. 
The elbow point of a curve is loosely defined as "the point of maximum
curvature"~\cite{salvador2004determining}, so there is no universally accepted
definition, but instead there are various heuristic approaches~\cite{satopaa2011finding, salvador2004determining, zambelli2016data, Antunes2018KneeElbowEB}
to find the point.
We use one variant of them, which defines an elbow point as the point near the
maximum amount of difference
(equal to \textit{Maximum Difference} in Zambelli et al.~\cite{zambelli2016data}).
Formally, we apply the elbow method on the \textit{mdist} curve: it stops at the number of
clusters $k$ right before the largest increase of minimum intercluster distance, defined as follows: 
\begin{equation}
\label{eq:stop}
k = argmax_{1 \leq k \leq N}\left(mdist_{k} - mdist_{k+1}\right)
\end{equation}
where $mdist_{1}$ and $mdist_{N+1}$ are set to $0$ and $1$, respectively,
assuming that we use a normalised distance metric.
For example, if this is applied to the $mdist$ curve in
Figure~\ref{fig:dendrogram_example}, the suggested number of clusters is two 
because the maximum difference is 0.55 ($= mdist_{2} - mdist_{3}$).
This method assumes that once the test cases are optimally clustered,
the increase in the minimum intercluster distance would be substantially high.

\section{Experimental Setup}
\label{sec:experimental_setup}
Using our Java and C multi-fault datasets, we compare the performance of our hypergraph-based method
to other approaches, including MSeer, in terms of the effectiveness and efficiency for failure
clustering and also the improvement in SBFL performance.

\subsection{Construction of multi-fault Dataset}

We construct multi-fault datasets using the existing
fault datasets \dfj~\cite{just2014defects4j} and Software-artifact Infrastructure Repository (SIR)~\cite{do2005supporting, SIR}.
Table~\ref{tab:subjects} shows the number of faulty subjects used in our experiment. As well as the multi-fault data (\# faults > 1),
we also include the single fault data (\# faults = 1) that have multiple failing test cases.
This is because we also need to check whether a clustering algorithm assigns multiple failing test
cases that share a single root cause to the same cluster, instead of dividing them into separate clusters.
Section~\ref{sec:experiment:java_faults} and Section~\ref{sec:experiment:c_faults}
explain the multi-fault data creation process for Java and C programs, respectively.
We measure the statement-level test coverage for all subjects
using \texttt{Cobertura} for Java and \texttt{gcov} for C.
Note that we make the failure-to-single-fault assumption, as in previous work on
failure clustering~\cite{podgurski1993partition, liu2006proximity, bowring2004active}:
we expect each of the  failing test cases to have a single root cause.

\begin{table}[t]
    \renewcommand{\arraystretch}{0.9}
    \caption{Experiment Dataset ($M$ = \# lines)}   %
    \label{tab:subjects}
    \centering
    \begin{tabular}{c|ccccccc|cc}
    \toprule
    \multirow{2}{*}{Base Dataset} & \multicolumn{7}{c|}{\# Faults} & \multirow{2}{*}{$M$} & \multirow{2}{*}{$|T_F|$}\\\cline{2-8}
                                   &   1 &   2 &  3 &  4 &  5 &  6 &  7 & &\\
    \midrule
    Defects4J (Java)               & 124 & 240 & 79 & 16 & 9 &  5 &  2 & 24069.6 & 4.8 \\
    SIR (C)                        &  54 &  54 & 19 & 3  & - &  - &  - & 1934.3 & 31.6 \\
    \midrule
    Total                          &  178 & 294 & 98 & 19 & 9 & 5 & 2 &  -  &   -  \\
    \bottomrule
    \end{tabular}
    \end{table}

\subsubsection{Java faults}
\label{sec:experiment:java_faults}

\dfj~\cite{just2014defects4j} is a real-world faults dataset from various open-sourced Java programs.
We use the five projects,
\textit{Lang}, \textit{Chart}, \textit{Time}, \textit{Math} and \textit{Closure},
to construct our multi-fault dataset. 

Each buggy version of the program in \dfj has a set of failing test cases
that reveals a single fault in the program.
After applying the provided revision patch on the buggy version, all fault-revealing test cases do not fail anymore.
All buggy versions of a project are sorted chronologically
by the date of revision, and the more recently a bug is fixed,
the lower the bug ID is assigned.
Therefore, in a project, it is likely that a buggy program with a higher bug
ID already contains the faults in the buggy programs with lower bug IDs,
which is not yet detected due to the absence of fault-revealing test cases for the faults.
For example, the more recently fixed faults, \texttt{Math-3} and \texttt{Math-4},
already exist in the older version \texttt{Math-5}.
We include such buggy versions that already contain multiple faults to our dataset.

\begin{figure}[t]
    \centering
    \includegraphics[width=0.6\textwidth]{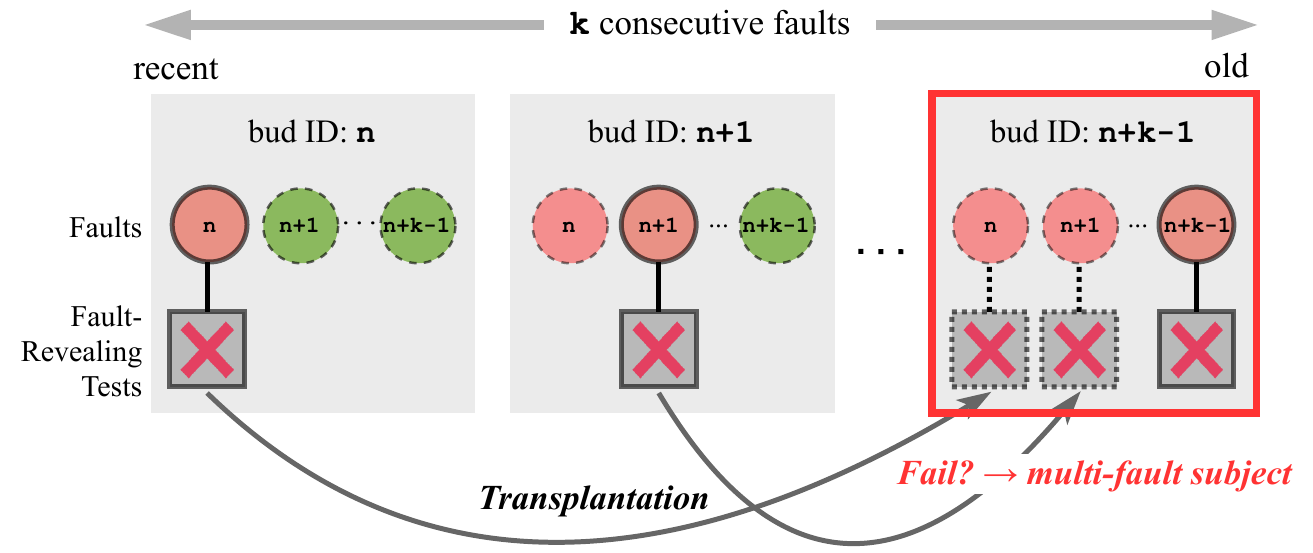}
    \caption{Building Java multi-fault dataset using \dfj}
    \label{fig:d4j_multifault}
\end{figure}

Since it is cumbersome to manually validate whether each fault exists in the older version,
we use the results of the fault-revealing test cases of the fault.
As shown in Figure~\ref{fig:d4j_multifault}, suppose that there are $k$ consecutive
buggy programs ($k>1$) whose fault-revealing test cases are \textit{non-overlapping} due to the failure-to-single-fault assumption.
For $i=0,\dots,k-2$,
we transplant the code snippets of fault-revealing test cases from the more recent buggy versions,
$\{n, \dots, n+k-2\}$, to the oldest buggy version $n+k-1$.
After the transplantation, we regard that a fault $n+i$ \textit{exists} in the buggy version $n+k-1$
if \textit{all} fault-revealing test cases of the fault are still compilable and fail on the version.

If all faults are considered to exist,
we add the constructed multi-fault subject (the buggy version $n+k-1$ with the $k$
sets of failing test cases) to our dataset.
Once we succeed to generate the $k$-faults subject from the buggy versions
$\{n, \dots, n+k-1\}$, we successively try to combine $k+1$ faults using the versions
$\{n-1, \dots, n+k-1\}$.

\subsubsection{C faults}
\label{sec:experiment:c_faults}
In addition to the Java fault dataset, we use four C subject programs from
SIR~\cite{do2005supporting, SIR} that is a widely-used debugging benchmark and
also employed in the previous failing clustering work~\cite{digiuseppe2012software, gao2017mseer}:
version 1.5 of \textit{gzip}, version 1.2 of \textit{grep}, version 1.1 of \textit{flex},
and version 2.0 of \textit{sed}.
The benchmark contains artificial faults seeded on a correct version of a program.
Each faulty region can be either activated or deactivated using macro
definitions and preprocessors, i.e., \texttt{ifdef}, or \texttt{ifndef}.
Note that a fault can span multiple (not necessarily continuous) lines, sharing the same root cause.
Because those faulty regions are distinct from each other,
multi-fault programs can be constructed by simultaneously activating
the multiple faulty lines. %
Then, we observe the failing tests of the combined faults. 

After combining the faults, in some cases, \textit{fault interference}~\cite{debroy2009insights, digiuseppe2011influence}
makes it difficult to clearly define the membership of some failing test cases.
For example, if a test case that is failing in the combined version was initially passed with any of the single faults,
it is hard to assign the cluster membership of the test case to only one fault since the presence of multiple faults makes the test fail.
Under the failure-to-single-fault assumption, we exclude such failing test cases from our evaluation dataset.
Similarly, only non-overlapping failing test cases of the faults are set to the target of clustering. 
We include only the multi-fault subjects of which each fault is not
entirely masked~\cite{debroy2009insights, jones2007debugging} by other faults;
therefore, all faults of the subjects can be discoverable by at least one failing test case.

Additionally, failing test cases which crashed due to illegal memory access (segmentation fault) are omitted 
since coverage data is not generated on those executions using \texttt{gcov}. 

\subsection{Evaluation Methodology}
\subsubsection{Clustering Performance}
When the ground-truth clustering is unknown,
the quality of clustering is typically evaluated by \textit{internal criteria}
such as the degree of cluster cohesion or separation.
However, satisfying the internal criteria does not always guarantee high effectiveness
in an application~\cite{cambridge2009online}.
In our multi-fault datasets, since we know which failing tests are fault-revealing tests of which fault, 
we could regard that information as \textit{ground-truth}.
Thereby, instead of internal criteria, we use \textit{external criteria} which directly
compare the clustering results with the ground-truth clusters.

In Section~\ref{sec:background}, we introduced two external criteria: \textit{Homogeneity} ($h$) 
and \textit{Completeness} ($m$)~\cite{rosenberg2007v}. Given ground-truth clusters $P_{GT} = \{T_1, \dots, T_c\}$
and arbitrary clusters $P = \{T'_1, \cdots, T'_k\}$ that are both partitions of $T_F$, the two
criteria are formally defined by:
\begin{align}
\label{eq:homogeneity} h(P, P_{GT}) = 1 - \frac{H(P_{GT}|P)}{H(P_{GT})}\\
\label{eq:completeness} m(P, P_{GT}) = 1 - \frac{H(P|P_{GT})}{H(P)}
\end{align}

In Eq.~\ref{eq:homogeneity}, $H(P_{GT}|P)$ means the \textit{conditional entropy} of the clusters $P_{GT}$ given the clusters $P$, and $H(P_{GT})$ is an \textit{entropy} of $P_{GT}$:
\begin{align*}
\small
H(P_{GT}|P) &= -\sum_{i=1}^{c}\sum_{j=1}^{k}\frac{|T_i \cap T'_j|}{|T_F|}log\frac{|T_i \cap T'_j|}{|T'_j|}\\
H(P_{GT}) &= -\sum_{i=1}^{c}\frac{|T_i|}{|T_F|}log\frac{|T_i|}{|T_F|}
\end{align*}
Similarly, $H(P|P_{GT})$ and $H(P)$ in Eq.~\ref{eq:completeness} are defined in a symmetric way.
Both $h$ and $m$ values are bounded in the range from 0 (worst) to 1 (best). 
A clustering is \textit{perfect} if and only if both $h$ and $m$ are 1.

A most widely-used external criterion in literature is \textit{normalised mutual information}
(NMI)~\cite{vinh2010information}, which measures the agreement between two clustering assignments.
Interestingly, it has been found that NMI is mathematically equivalent to the harmonic mean of Homogeneity and Completeness, i.e., $\frac{2(h+m)}{h \cdot m}$ (the proof is in \cite{becker2011identification}).
Therefore, we use NMI to evaluate failure clustering effectiveness.

\subsubsection{Fault Localisation Performance}
\label{sec:experiment:evaluation:fl}

Once we cluster the multiple failing test cases,
following the parallel debugging process~\cite{jones2007debugging}, we generate a
suspiciousness ranking from each cluster using the failing test cases in the cluster
along with all passing test cases.
The $k$ rankings obtained from $k$ failure clusters can be investigated by multiple developers in a parallel manner,
assigning the ranking from each cluster to a developer who is most responsible for the cluster's failing test cases. 
Once developers finish inspecting all rankings
and fix the found faults, the next iteration can begin with remaining failing test cases.
In this work, we evaluate the fault localisation performance of the first iteration
of parallel debugging.

\begin{figure}[ht]
\centering
\centerline{\includegraphics[width=0.7\textwidth]{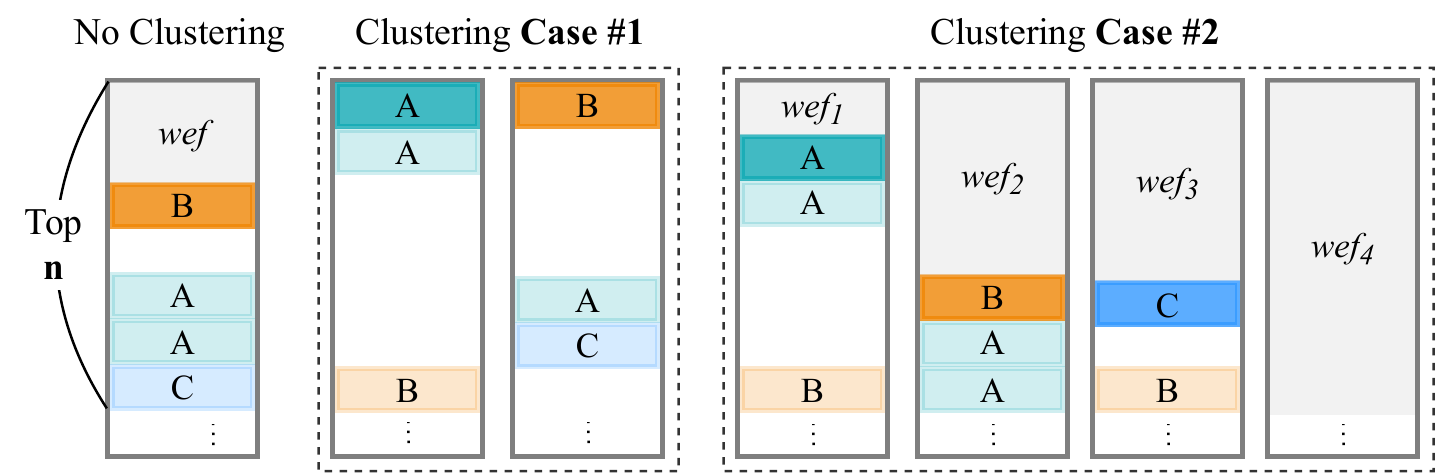}}
\caption{The example of the parallel debugging process}
\label{fig:fl_eval}
\end{figure}
Given a ranking, let us assume that developers inspect only the program components within the top $n$ (complying with the guideline from Parnin and Orso ~\cite{parnin2011automated})
until finding the highest-ranked faulty component.
A fault is considered to be \textit{found} when it is associated with at least one of the highest-ranked faulty components of the rankings.
For example, in Figure~\ref{fig:fl_eval}, two faults, $\{A, B\}$, and three faults, $\{A, B, C\}$, are found in Case \#1 and  \#2, respectively. Note that a fault may span multiple components, e.g., consecutive lines.

Wasted effort ($\textit{wef}_i$) for the $i$-th ranking is defined as the number of program components should be examined before finding the highest-ranked faulty component in the ranking (Figure~\ref{fig:fl_eval}). If no faulty component exists within top $n$, $\textit{wef}_i$ is $n$. Then, the total wasted effort (\textit{t-wef}) is defined as the sum of $\textit{wef}_i$ for all rankings: $\textit{t-wef} = \sum_{i=1}^{k}\textit{wef}_i$. For example, \textit{t-wef} of Case \#1 is less than the one of Case \#2 in Figure~\ref{fig:fl_eval}. A lower \textit{t-wef} means better efficiency of fault localisation.
We also compute the percentage of rankings that cannot rank at least one faulty element higher than other rankings. We call such rankings as \textit{redundant} rankings.

We use two FL techniques, Ochiai~\cite{abreu2009practical} and Crosstab~\cite{wong2008crosstab}.
Since our Java subjects have much more lines than C subjects,
we use the method-level rankings while the suspiciousness score of each method is set to the
maximum suspiciousness of its lines.

\subsection{Other Clustering Methods for Comparison}
\label{sec:experiment:other_methods}

\name is compared with following failure clustering methods:
\begin{itemize}
    \item MSeer~\cite{gao2017mseer}: A recently proposed failure clustering technique using the
    K-medoids algorithm with own technique for estimating $K$.
    Revised Kendall-Tau (RKT) distance is used to calculate the distance between failing test cases.
    \item Test Class Name (TCN): Assigning the cluster membership of failing test case methods according to the classes of them (only for Java subjects)
    \item Agglomerative Hierarchical Clustering (AHC) with other distance metrics: Jaccard, S{\o}rensen-Dice, Cosine, Euclidean, Hamming, and RKT (used in MSeer).
    \footnote{When representing a failing test as a set or a vector, we consider only the components covered by at least one failing test. RKT is min-max normalised for each subject.}
\end{itemize}

\section{Result and Analysis}
\label{sec:results}

To evaluate \name, we set up the following research questions:
\begin{itemize}
\item \textbf{RQ1. Distance Metric}: How effective is \textit{hdist} when compared to other
distance metrics in failure clustering?
\item \textbf{RQ2. Stopping Criteria}: How accurate is the failure clustering with our stopping
criterion when compared to other clustering approaches?
\item \textbf{RQ3. Efficiency}: How efficient is the calculation of \textit{hdist} when compared to other distance metrics?
\item \textbf{RQ4. FL accuracy}: How does the accuracy of failure clustering affect SBFL performance?
\end{itemize}

In the following sections, we present answers to our research questions.
Full results are available at our repository.\footnote{\resultlink}

\subsection{\textbf{RQ1: Distance Metric}}

\begin{figure}[t]
\begin{subfigure}[b]{0.3\textwidth}
    \centering
    \includegraphics[width=\textwidth]{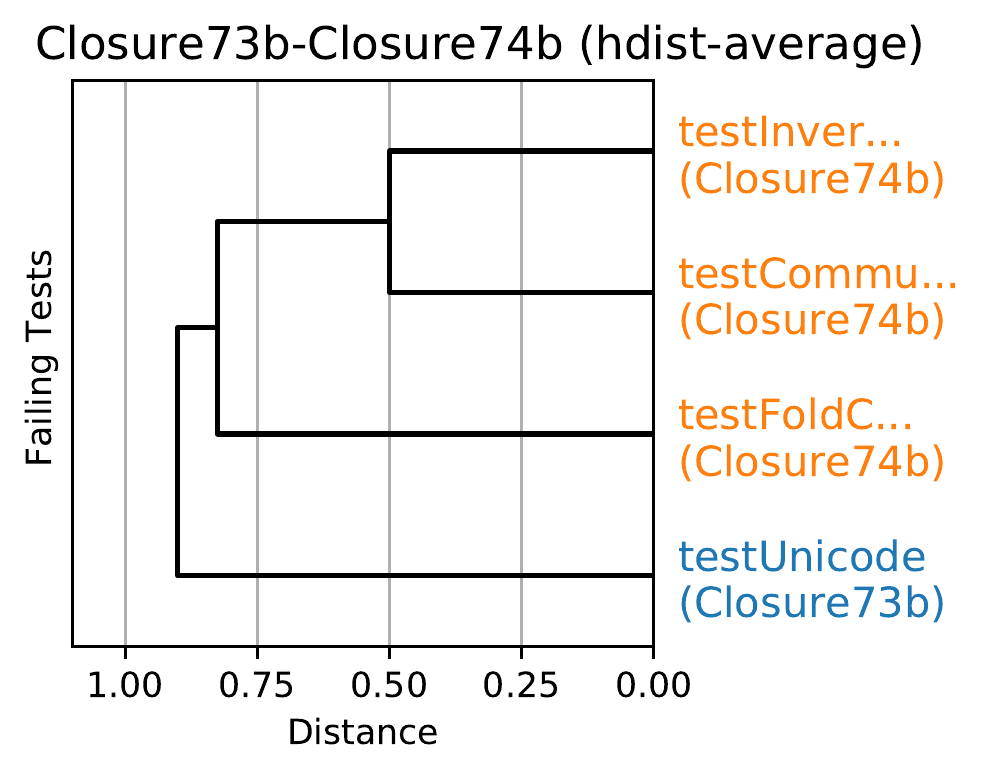}
    \caption{Using \textit{hdist}-average}
    \label{fig:dendrogram_hdist}
\end{subfigure}
\begin{subfigure}[b]{0.3\textwidth}
    \centering
    \includegraphics[width=\textwidth]{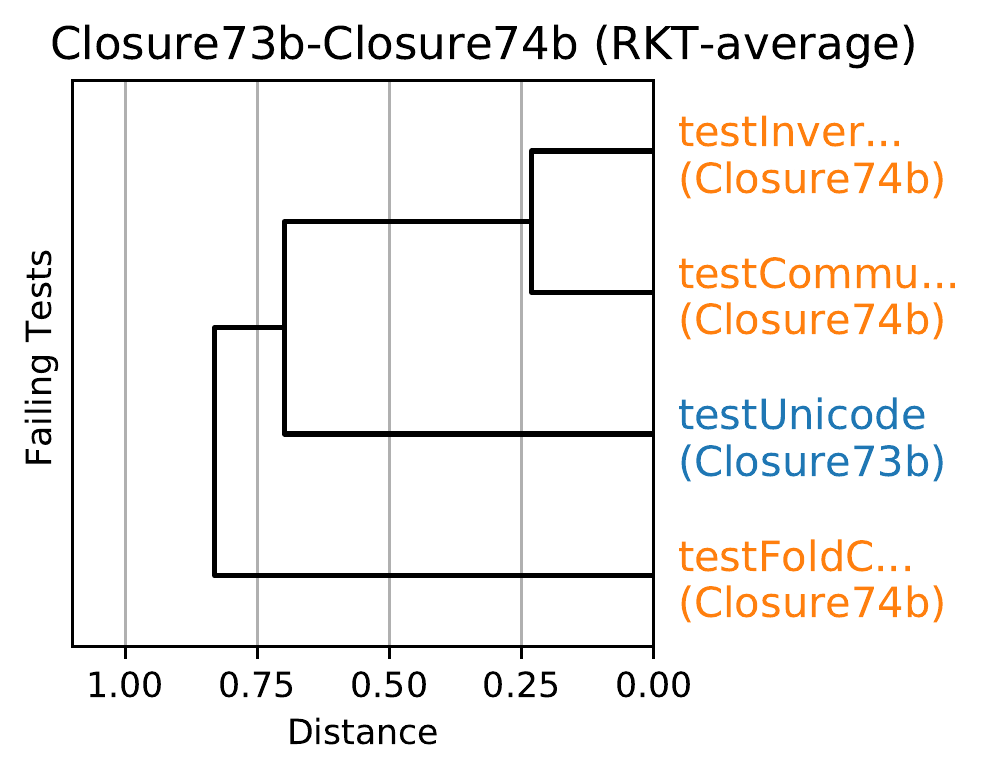}
    \caption{Using RKT-average}
    \label{fig:dendrogram_RKT}
\end{subfigure}
\caption{Dendrogram for Closure73b-Closure74b}
\label{fig:Closure73b-Closure74b}
\end{figure}

\begin{figure}[t]
    \begin{subfigure}[b]{0.6\textwidth}
        \centering
        \includegraphics[width=\textwidth]{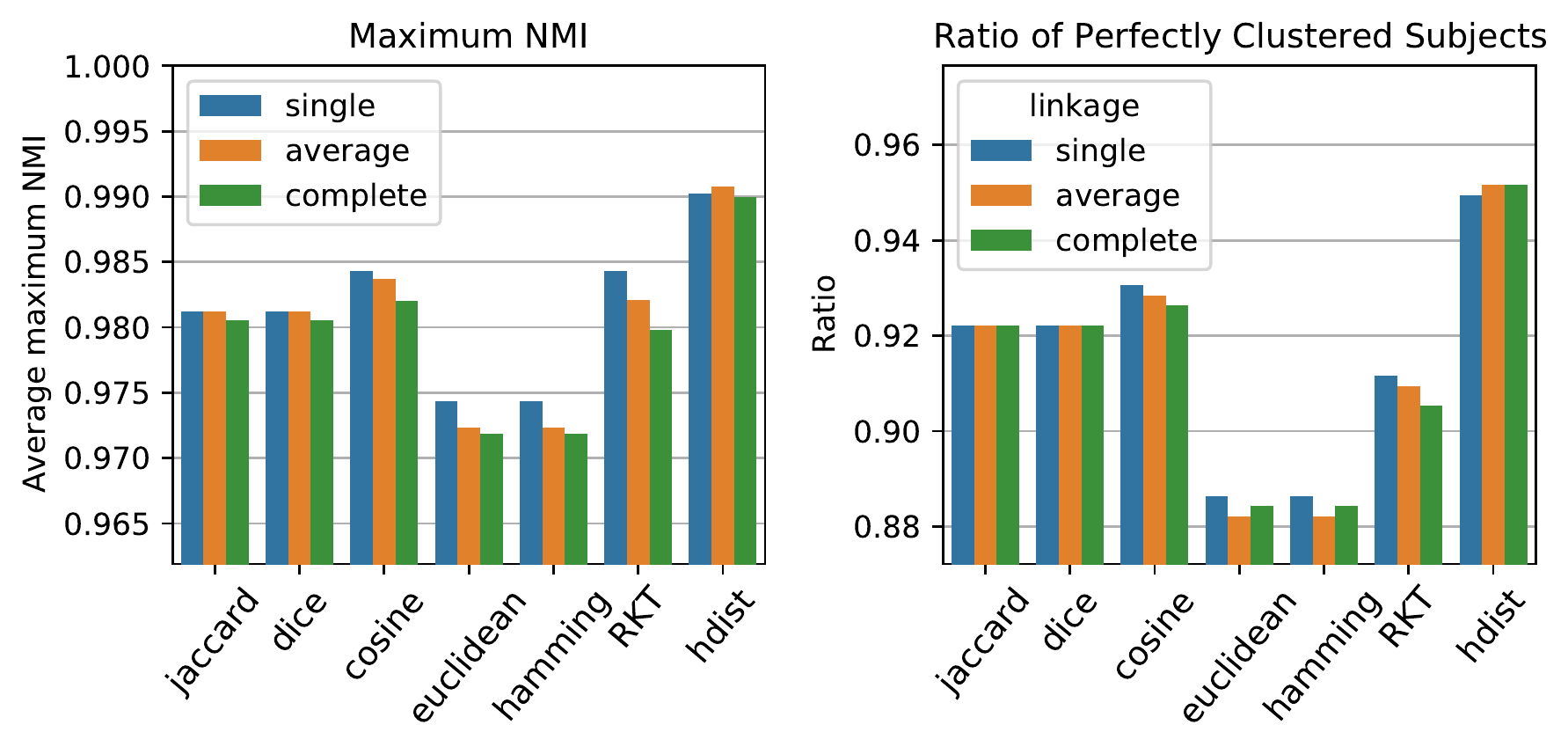}
        \caption{On 475 Java subjects}
        \label{fig:optimal_Java}
    \end{subfigure}
    \begin{subfigure}[b]{0.6\textwidth}
        \centering
        \includegraphics[width=\textwidth]{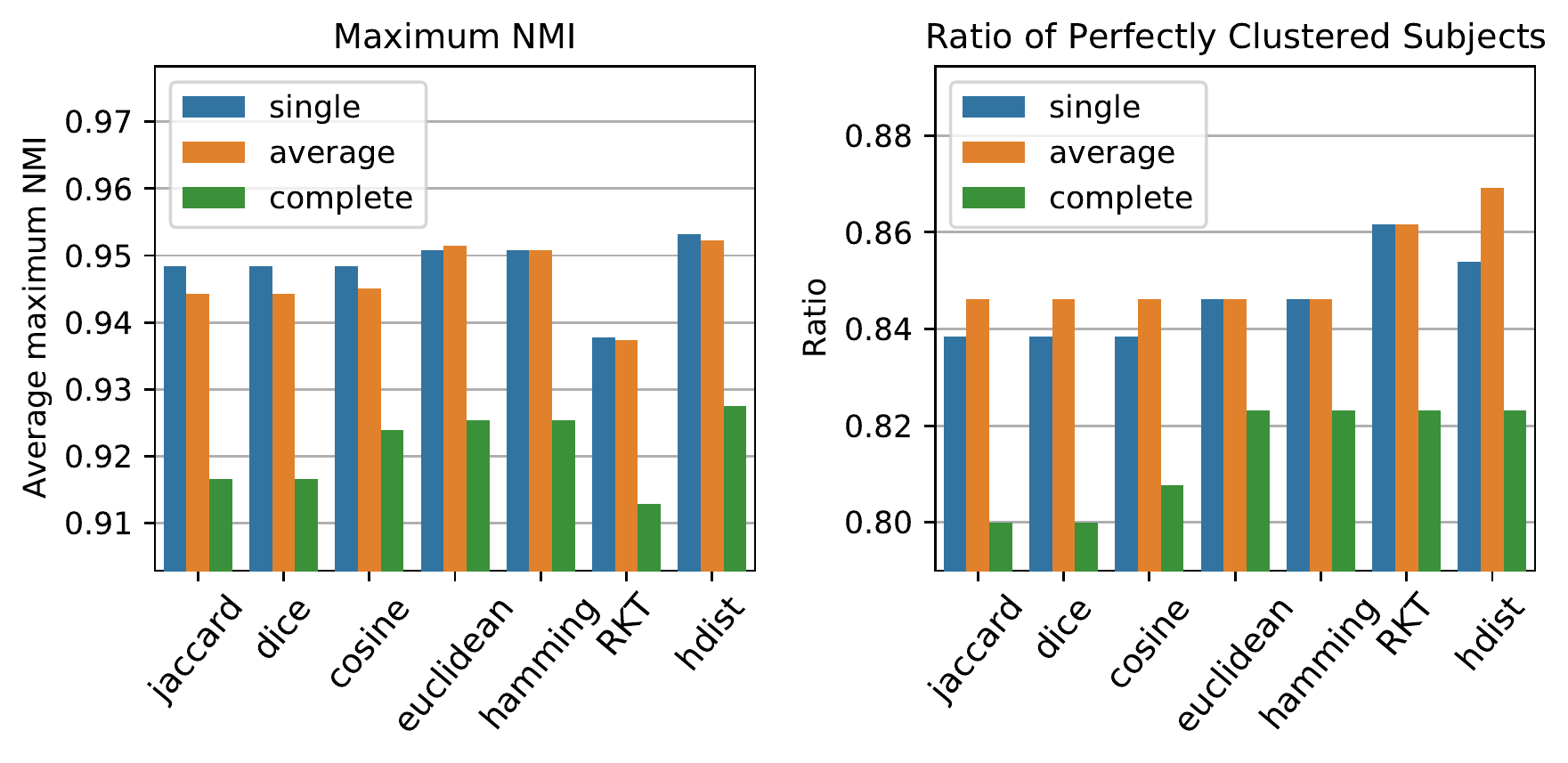}
        \caption{On 130 C subjects}
        \label{fig:optimal_C}
    \end{subfigure}
    \caption{Averaged maximum NMI (left) and the ratio of perfectly clustered subjects (right). Note that the ranges of the y-axis are different.}
    \label{fig:optimal}
    \end{figure}

To answer RQ1, we compute the maximum NMI values among all iterations of AHC 
without considering stopping criteria. In addition, we check whether the 
failing test cases of each faulty subject are perfectly clustered at any 
stopping point of AHC. For example, in Figure~\ref{fig:Closure73b-Closure74b},
the maximum NMI of \textit{hdist} is $1$, as failing tests
are perfectly clustered after two merges (\# clusters = 2). In comparison, the 
maximum NMI of RKT is $0.70$ (\# clusters = 3), and the failing tests are 
not perfectly clustered with any number of clusters.

Figure~\ref{fig:optimal} shows the maximum NMI values averaged over all subjects (left) and the
ratio of perfectly clustered subjects (right). This can be regarded as a performance measure of
failure clustering with an \textit{optimal} stopping criterion for each distance metric.
Note that 34.5\% of Java subjects and 9.2\% of C subjects have only two failing test 
cases, always resulting in perfect clustering (e.g., NMI = 1).
Excluding those subjects,
on Java, \textit{hdist}-average outperforms other distance
metrics in terms of NMI;
NMI values of \textit{hdist}-average are significantly higher than
ones of RKT-single (dependent $t$-test for paired samples with p=0.023).
On C, \textit{hdist}-single and -average show higher mean values when
compared to other distance metrics, but the differences are not statistically significant.
In terms of the perfectly clustered ratio, \textit{hdist}-average
shows the best performance: it perfectly clusters about 95\% and 87\% of Java 
and C subjects, respectively, in one of the stopping points in AHC. 

Interestingly, we observe that 
RKT does not always outperform set- or vector-based distance metrics, 
which is inconsistent with the core idea of Gao and Wong~\cite{gao2017mseer} 
that assumes that ranking-based distance metrics will outperform other
distance metrics.

\noindent \textbf{Answer to RQ1}: \textit{hdist} (especially with average 
linkage) outperforms other distance metrics for failure clustering.

\subsection{\textbf{RQ2: Stopping Criteria}}
\begin{table}[t]
\renewcommand{\arraystretch}{1.0}
\caption{Failure clustering performance using Eq.~\ref{eq:stop} for AHC
(\textit{k} = \# clusters, \textit{c} = \# actual faults, \textit{h} = homogeneity,
\textit{m} = completeness, Perf. = the ratio of perfectly clustered subjects)}
\label{tab:RQ2}
\centering
\begin{tabular}{p{0.65cm}|c|rp{0.43cm}p{0.43cm}p{0.43cm}r|rp{0.43cm}p{0.43cm}p{0.43cm}r}
\toprule
\multicolumn{2}{c|}{\multirow{2}{*}{\makecell[c]{Clustering\\Method}}}    & \multicolumn{5}{c|}{\textbf{Java}} & \multicolumn{5}{c}{\textbf{C}}\\\cline{3-12}
\multicolumn{2}{c|}{} & $k/c$ &\textit{h} & \textit{m} & \textbf{NMI} & \textbf{Perf.} & $k/c$ & \textit{h} & \textit{m} & \textbf{NMI} & \textbf{Perf.}\\
\midrule
\multicolumn{2}{c|}{MSeer}                     & 1.67 & 0.985 & 0.601 & 0.644 & 0.406 & 3.19 & 0.855 & 0.479 & 0.427 & 0.246 \\\hline
\multicolumn{2}{c|}{TCN}                       & 1.26 & 0.950 & 0.818 & 0.802 & 0.646 & - & - & - & - & -\\\hline
           
\multirow{2}{*}{Jacc.}          & min          & 1.02 & 0.826 & 0.907 & 0.753          & 0.674          & 0.86 & 0.688 & 0.967 & 0.670          & 0.585          \\
                                & avg          & 1.03 & 0.830 & 0.901 & 0.753          & 0.669          & 0.88 & 0.694 & 0.959 & 0.669          & 0.585          \\
                                & max          & 1.04 & 0.840 & 0.894 & 0.761          & 0.674          & 0.90 & 0.728 & 0.952 & 0.693          & 0.615          \\ \cline{1-12}
\multirow{2}{*}{Dice}           & min          & 0.90 & 0.717 & 0.953 & 0.685          & 0.621          & 0.79 & 0.616 & 0.993 & 0.620          & 0.554          \\
                                & avg          & 0.92 & 0.721 & 0.946 & 0.684          & 0.619          & 0.82 & 0.645 & 0.974 & 0.638          & 0.562          \\
                                & max          & 0.94 & 0.729 & 0.935 & 0.686          & 0.615          & 0.83 & 0.654 & 0.967 & 0.633          & 0.577          \\ \cline{1-12}
\multirow{2}{*}{Cos.}           & min          & 0.88 & 0.701 & 0.962 & 0.676          & 0.617          & 0.77 & 0.569 & 0.987 & 0.566          & 0.508          \\
                                & avg          & 0.90 & 0.702 & 0.957 & 0.673          & 0.615          & 0.78 & 0.578 & 0.978 & 0.567          & 0.523          \\
                                & max          & 0.91 & 0.709 & 0.950 & 0.675          & 0.613          & 0.81 & 0.622 & 0.968 & 0.602          & 0.562          \\ \cline{1-12}
\multirow{2}{*}{Ham.}           & min          & 0.88 & 0.690 & 0.926 & 0.641          & 0.560          & 0.76 & 0.578 & 0.992 & 0.582          & 0.508          \\
                                & avg          & 0.89 & 0.704 & 0.916 & 0.658          & 0.564          & 0.80 & 0.615 & 0.980 & 0.613          & 0.531          \\
                                & max          & 0.93 & 0.746 & 0.899 & 0.698          & 0.577          & 0.82 & 0.650 & 0.955 & 0.637          & 0.531          \\ \cline{1-12}
\multirow{2}{*}{RKT}            & min          & 1.36 & 0.957 & 0.705 & 0.697          & 0.539          & 1.26 & 0.831 & 0.802 & 0.646          & 0.577          \\
                                & avg          & 1.42 & 0.971 & 0.689 & 0.695          & 0.531          & 1.82 & 0.900 & 0.710 & 0.627          & 0.538          \\
                                & max          & 1.42 & 0.976 & 0.673 & 0.686          & 0.524          & 2.39 & 0.938 & 0.561 & 0.530          & 0.408          \\ \cline{1-12}
\multirow{2}{*}{\textbf{hdist}} & min          & 1.20 & 0.949 & 0.848 & 0.826          & 0.678          & 0.86 & 0.718 & 0.989 & 0.713          & 0.677          \\
                                & \textbf{avg} & 1.21 & 0.958 & 0.846 & \textbf{0.833} & \textbf{0.680} & 0.94 & 0.798 & 0.971 & \textbf{0.776} & \textbf{0.731} \\
                                & max          & 1.22 & 0.963 & 0.839 & 0.832          & \textbf{0.680} & 1.35 & 0.846 & 0.835 & 0.709          & 0.585          \\
\bottomrule
\end{tabular}
\end{table}

For AHC, we determine the number of clusters, $k$, using our stopping criterion defined in Eq.~\ref{eq:stop}.
Table~\ref{tab:RQ2} shows the failure clustering performance of AHC and the other clustering methods,
MSeer and TCN, described in Section~\ref{sec:experiment:other_methods}. The second column shows the
method of defining the intercluster distance. Euclidean distance is excluded since it is not normalised.

With our stopping criterion, AHC with \textit{hdist}-average outperforms other
clustering approaches on both Java and C datasets in terms of NMI and the ratio of perfectly
clustered subjects (\textit{Perf.}). It perfectly
maps failures to their root causes for 68\% and 73\% of the studied Java and C subjects, respectively.
Since RQ1 shows that the upper-bounds of \textit{Perf.} are 95\% and 87\%, there is still room for
improvement of a stopping criterion. We have evaluated the other two stopping criteria:
stopping after reaching some distance threshold (\textit{threshold-based}) and at maximum modularity
of pairwise distance graph~\cite{blondel2008fast} (\textit{modularity-based}).
Briefly, the modularity-based criterion shows poorer performance than our
stopping criterion for every distance metric. Meanwhile, the threshold-based criterion shows a
discrepancy between best-performing thresholds on Java and C subjects, even though their NMI scores
are higher than our stopping criterion: \textit{hdist}-single with the threshold 0.65 shows the best
NMI score, 0.86, on Java, while \textit{hdist}-single with the threshold 0.45 shows the best
NMI score, 0.79, on C. We need further research to understand the features of faulty programs
that could be used to determine a good distance threshold.

On both Java and C subjects, MSeer tends to generate many more clusters than the actual number of faults.
The clusters of MSeer show relatively high homogeneity, but lower completeness than all others, which
means the failing tests due to the same fault is likely to split into different clusters.
Interestingly, AHC-RKT with our stopping criterion outperforms MSeer that also uses RKT.
We note that TCN performs relatively
well for Java. Since related tests are likely to be put in the same class, TCN can be viewed as
"manual clustering" by developers. 
However, we expect its performance to depend heavily on the organisation of the test classes.

\noindent \textbf{Answer to RQ2}: Using our stopping criterion, especially with \textit{hdist}-average, outperforms the TCN baseline and the existing failure clustering approach, MSeer.

\subsection{\textbf{RQ3: Efficiency}}
\label{sec:result:RQ3}

\begin{figure}[t]
    \centering
    \includegraphics[width=0.6\textwidth]{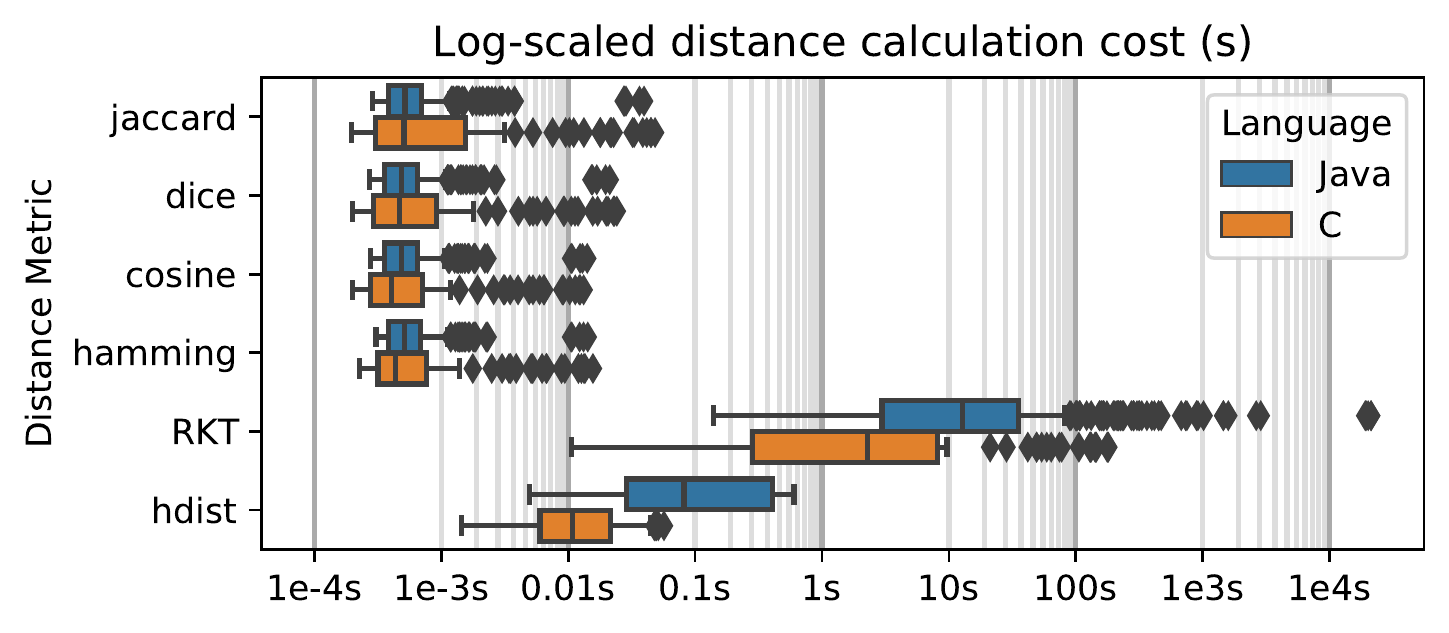}
    \caption{Distribution of distance calculation cost (log-scale)}
    \label{fig:efficiency}
\end{figure}

\begin{figure}[t]
    \centering
    \includegraphics[width=0.8\textwidth]{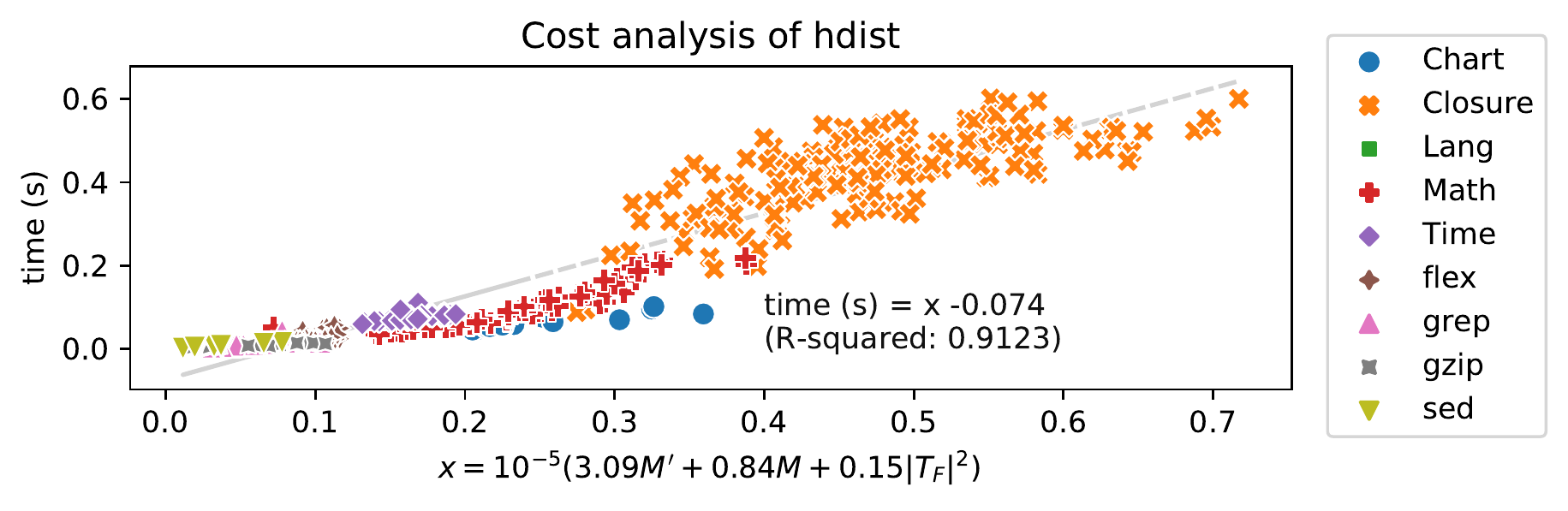}
    \caption{Analysis of \textit{hdist} calculation time (s)}
    \label{fig:hdist_complexity}
\end{figure}

Figure~\ref{fig:efficiency} shows the log-scaled distribution of distance calculation time
on all studied subjects. \footnote{Measured on a PC with Intel Core i7-7700 CPU and 32GB memory.}
Note that the cost of \textit{hdist} includes not only the distance computation but also the
hypergraph modelling and the subgraph extraction process.
For every subject, calculating \textit{hdist} requires one second at most.
In Figure~\ref{fig:hdist_complexity}, we present the cost analysis of \textit{hdist} using linear regression.
The required computation time is approximately linear to $M$, $M'$ and $|T_F|^2$:
$\textit{time} (s) = 10^{-5}\times(3.09M'+0.84M'+0.15|T_F|^2)$ ($R^2 = 0.9123$),
where $M$ is total number of program components, and $M'$ is the number of program components
executed by at least one failing test case, i.e., the number of hyperedges in $G_{T_F}$
(Eq.~\ref{eq:restriction}). 

The computation of RKT, on the other hand, remains expensive despite efforts 
for optimisation: it can takes more than 10,000 seconds (i.e., more than 2.7 
hours) to calculate distances between failing tests for some of the large 
subjects. For example, it takes 5.6 hours (2.5 hours with GPU parallelisation) 
to calculate RKT distances between failing tests of \texttt{Closure49b-50b} 
that has 38,235 executed lines and 68 failing test cases. In comparison, 
\textit{hdist} only takes 0.49 seconds. Unlike \textit{hdist} that is roughly linear to $M$, RKT shows $O(M^2|T_F|^2)$ time complexity.\footnote{Linear regression for RKT:$\textit{time} (s) = 3.0 \times 10^{-9} \times M^2|T_F|^2 - 2.7$ ($R^2 = 0.9996$)}

\noindent \textbf{Answer to RQ3}: Even though the cost of \textit{hdist} calculation
is higher than the set- or vector-based metrics, it is still negligible even in the large subjects,
while RKT requires much more computation time.

\begin{table}[t]
    \caption{FL performance after the failure clustering (using Ochiai, max tie-breaker). \textit{R.R.} means Redundant Rankings.}
    \label{tab:RQ4}
    \centering
    \begin{tabular}{c|c|c|r|cccc}
    \toprule
    \multicolumn{3}{c|}{Method}  & \textit{R.R.} & \multicolumn{4}{c}{The ratio of found faults (\textit{t-wef})}\\\hline\hline
    \multicolumn{4}{c|}{\textbf{Java (method-level)}} & n=1 & n=5 & n=10 & n=$\infty$\\\hline
    \multicolumn{3}{c|}{No Clustering} & - & 0.13 (0.73) & 0.34 (2.56) & 0.38 (3.88) & 0.50 (21.6)\\\hline
    \multicolumn{3}{c|}{MSeer} & 19.2\% & 0.37 (2.11) & 0.75 (7.07) & 0.82 (11.19) & 0.98 (285.9)\\\hline
    \multicolumn{3}{c|}{TCN} & 9.2\% & 0.33 (1.71) & 0.73 (5.48) & 0.80 (8.62) & 0.97 (234.1)\\\hline
    & \multirow{3}{*}{Jacc.} & min & 4.2\% & 0.31 (1.34) & 0.68 (4.16) & 0.74 (6.36) & 0.91 (162.3)\\
    &  & avg & 4.4\% & 0.31 (1.36) & 0.68 (4.22) & 0.74 (6.48) & 0.91 (177.2)\\
    &  & max & 4.6\% & 0.31 (1.38) & 0.68 (4.28) & 0.74 (6.55) & 0.92 (191.9)\\\cline{2-8}
    A& \multirow{3}{*}{RKT} & min & 12.8\% & 0.35 (1.72) & 0.74 (5.50) & 0.80 (8.53) & 0.97 (241.5)\\
    H&  & avg & 13.2\% & 0.35 (1.77) & 0.75 (5.61) & 0.81 (8.67) & 0.97 (249.9)\\
    C&  & max & 14.0\% & 0.35 (1.78) & 0.75 (5.70) & 0.81 (8.81) & 0.98 (250.6)\\\cline{2-8}
    & \multirow{3}{*}{hdist} & min & 6.0\% & 0.35 (1.56) & 0.73 (4.86) & 0.79 (7.45) & 0.97 (181.2)\\
    &  & avg & 6.7\% & 0.35 (1.59) & 0.73 (4.99) & 0.80 (7.70) & 0.97 (234.1)\\
    &  & max & 6.3\% & 0.36 (1.60) & 0.74 (5.02) & 0.80 (7.80) & 0.98 (235.6)\\\hline\hline
    \multicolumn{4}{c|}{\textbf{C (line-level)}} & n=5 & n=10 & n=15 & n=$\infty$\\\hline
    \multicolumn{3}{c|}{No Clustering} & - & 0.05 (4.8) & 0.09 (9.0) & 0.12 (13.1) & 0.73 (342.5)\\\hline
    \multicolumn{3}{c|}{MSeer} & 41.1\% & 0.06 (22.2) & 0.14 (43.0) & 0.17 (63.2) & 0.91 (2698.1)\\\hline
    & \multirow{3}{*}{Jacc.} & min & 3.5\% & 0.05 (7.1) & 0.11 (13.6) & 0.15 (19.7) & 0.87 (444.5)\\
    &  & avg & 4.3\% & 0.05 (7.2) & 0.11 (13.8) & 0.15 (20.0) & 0.87 (467.6)\\
    &  & max & 4.8\% & 0.06 (7.7) & 0.12 (14.7) & 0.16 (21.1) & 0.89 (462.1)\\\cline{2-8}
    A& \multirow{3}{*}{RKT} & min & 14.7\% & 0.05 (10.5) & 0.13 (20.2) & 0.17 (29.6) & 0.94 (706.8)\\
    H&  & avg & 22.6\% & 0.04 (15.0) & 0.13 (29.3) & 0.16 (43.1) & 0.96 (1747.0)\\
    C&  & max & 32.6\% & 0.04 (18.3) & 0.13 (35.9) & 0.16 (52.8) & 0.98 (2137.1)\\\cline{2-8}
    & \multirow{3}{*}{hdist} & min & 2.3\% & 0.06 (7.3) & 0.13 (13.9) & 0.17 (20.1) & 0.89 (480.8)\\
    &  & avg & 4.4\% & 0.05 (8.2) & 0.13 (15.6) & 0.17 (22.5) & 0.93 (508.1)\\
    &  & max & 15.0\% & 0.05 (11.7) & 0.13 (22.3) & 0.17 (32.1) & 0.96 (653.6)\\
    \bottomrule
    \end{tabular}
    \end{table}

\subsection{RQ4: FL Accuracy}

Table~\ref{tab:RQ4} shows the FL performance using Ochiai after the failure clustering.
The lowest ranks are assigned to the program components with equal suspiciousness scores (\textit{max} tie-breaker).
We do not present the results of Dice, Cosine, and Hamming because using
each of them found fewer faults than using Jaccard.\footnote{The results using other distance metrics
(Dice, Cosine, and Hamming), FL technique (Crosstab), and tie-breakers (min) are available in our
experiment repository.} Note that Eq.~\ref{eq:stop} is used as a
stopping criterion for AHC.

The results show that we can find more faults with failure clustering.
Especially, MSeer, AHC-RKT,
and AHC-\textit{hdist} similarly find the greatest number of faults: when $n=10$, about 80-82\% and
13-14\% of faults can be found on Java and C subjects, respectively. In RQ2, we show that MSeer and
AHC-RKT produce more homogeneous but less complete clusters than AHC-\textit{hdist}.
Although there is less noise when performing FL due to the
higher homogeneity, fewer failing test cases of a fault are utilised in FL due to the lower completeness.
This paucity of information may degrade FL accuracy so that a similar number of faults being found
by MSeer and AHC-RKT when compared to AHC-\textit{hdist} despite their higher homogeneity.
Meanwhile, the lower completeness also make MSeer and AHC-RKT produce more redundant rankings than
AHC-\textit{hdist}. This, in turn, leads to a higher inspection cost, \textit{t-wef}. In fact,
AHC-\textit{hdist} (with the single linkage method) requires only 63\% and 18\% of \textit{t-wef}
than MSeer (when $n=\infty$) on Java and C subjects, respectively.
Consequently, AHC-\textit{hdist} shows higher efficiency than the other methods that found a similar
number of faults. 

\noindent \textbf{Answer to RQ4}: The more homogeneous a cluster is, the more effective the FL is. Similarly, the more complete a cluster is, the more efficient the FL is (as pointed out in Section~\ref{sec:background}).

\section{Threats to Validity}
\label{sec:threats}

Threats to internal validity concern factors that may influence the observed 
effects, such as the integrity of the coverage and test result data, as well 
as failure clustering and fault localisation. To minimise
threats, we use \texttt{Cobertura} and \texttt{gcov}, both widely used 
coverage profilers. We also make both our implementation and datasets 
publicly available for further scrutiny.

Threats to external validity concern any factor that may limit the 
generalisation of our results. Our results are based on \dfj and SIR: both 
have been widely studied in conjunction with automated debugging techniques, 
and facilitate a direct comparison between our results and other existing work. 
However, only further studies with more diverse programs and faults can 
strengthen claims for generalisation. Additionally, the reported 
results are strictly based on the failure-to-single-fault 
assumption~\cite{digiuseppe2012software}, and may not generalise for 
multiple-fault failures~\cite{yu2015does}. We will consider overlapping 
clustering~\cite{xie2013overlapping, whang2018non, yang2013overlapping, whang-tkde2016}
to handle such cases in the future.
Our findings about \textit{hdist} are based on hierarchical clustering, and may 
not generalise to partitional clustering algorithms. Finally, our results are 
based on coverage measured at the statement level: the findings may not 
generalise to coverage measured at other granularity levels. We expect 
coarser-grained coverage criteria such as method or file coverage to perform 
worse, as they contain less information about the test behaviour. Since the 
statement coverage is one of the most widely used type of coverage in
practice, we believe the current experimental design can provide the most 
accurate evaluation of the proposed distance metric that is also practically 
relevant. A more thorough evaluation of the use of coarser-grained coverage 
metrics in situations where statement level coverage is not available will 
allow us to evaluate how much the clustering accuracy degrades due to the 
change of granularity. We leave this as future work.

Finally, threats to construct validity concern the use of metrics that may not reflect the 
properties we intend to measure. All evaluation metrics for clustering (homogeneity, 
completeness, and NMI) are widely used in the literature, leaving little 
room for misunderstanding. Following Parnin and Orso~\cite{parnin2011automated}, 
we report count based metrics for FL, under 
the assumption that developers will inspect only the first few  elements in 
the suspiciousness ranking.

\section{Related Work}
\label{sec:related_work}

We present related work in failure clustering for debugging and hypergraph clustering algorithms.

\subsection{Coverage-based Failure Clustering}
\label{sec:related_work:failure_clustering}
Podgurski et al. ~\cite{podgurski2003automated} showed
that the coverage profile could be used to group failures that share root causes.
Jones et al.~\cite{jones2007debugging} defined a parallel-debugging process,
which produces \textit{fault-focused} clusters of failing test cases to parallelise the task of
debugging faults.
They proposed two techniques for clustering, either representing a failing test as a behaviour model
or a suspiciousness ranking. The second technique, often used for baseline comparison
in several later work~\cite{steimann2012improving,gao2017mseer}, employs Tarantula~\cite{jones2002visualization}
with each failing test and all passing tests to generate rankings, and measure the similarity between them using Jaccard.
Golagha et al.~\cite{golagha2017reducing} adapt the parallel-debugging process into a real context with a high
cost of test execution, and suggest using only a representative test from each cluster rather than utilising
all failing tests for debugging. They do not evaluate automated FL
performance after failure clustering.

Recently, Gao and Wong proposed MSeer~\cite{gao2017mseer}, which defines
the profile of a failing test case as a suspiciousness ranking in a similar manner with
Jones et al.~\cite{jones2007debugging}.
They use a revised Kendall-Tau distance that performs better than Jaccard distance in comparing rankings
and employ the K-medoids clustering algorithm, using their own method of estimating the number of clusters. 

Some work \cite{steimann2012improving, hogerle2014more} suggests a partitioning method that divides
both failing tests and program elements using algorithms borrowed from integer linear programming. They only
account for failing test cases, thus not considering the differing importance of program components as
mentioned in Section~\ref{sec:distance_metrics}.

\subsection{Non Coverage-based Failure Clustering}
There are several existing work in clustering crashes or failing tests~\cite{dang2012rebucket, pham2017bucketing, van2018semantic, golagha2019failure}
without coverage information.
Dang et al.~\cite{dang2012rebucket} proposed ReBucket that clusters duplicate \textit{crash reports} using
call stacks. They designed a novel metric called \textit{Position Dependent Model}
to measure the similarity between two crash call stacks. However, this method can be applied to
only crash failures, and not to other failures such as assertion violations associated with test oracles.
Pham et al.~\cite{pham2017bucketing} proposed a symbolic execution-based failure clustering method.
Instead of clustering crash or bug reports, it clusters the failing test cases generated
during the symbolic execution while assuming no provided bug-revealing test input.
Since their approach is plugged into the main loop of a symbolic execution engine,
it cannot be easily applied to general test cases written by developers.
Tonder et al.~\cite{van2018semantic} presented an approach to bucket crashing inputs produced by
fuzzers. It transforms a program under test to characterise crashes based on the idea that
a correct fix can group crashing inputs triggered by the same, unique bug.
They evaluated their approach on six real-world projects with three different fuzzers.
Although utilising semantic analysis of a failure for failure clustering can yield precise results,
this approach also requires frequent recompiling and execution
of the program under test to validate fixes, which may limist its scalability to industry-level projects. In contrast, our distance metric, \hdist, and failure clustering technique, \name, only require the coverage information and do not incur any other expensive analysis cost. 
Golagha et al.~\cite{golagha2019failure} used non-code features to cluster failing test cases.
They cluster the test failure utilising general test features such as identifiers,
component membership, history data of test execution, broken/repaired 
features, or data collected
from the associated issue tracker such as Jira.
While this method does not require the cost of coverage measurement,
it cannot be applied to projects without sufficient issue tracking history, 
limiting its applicability to mature projects.

\subsection{Hypergraph Clustering}
\label{sec:related_work:hypergraph}
Hypergraph clustering is considered an important problem in the fields of data mining and
machine learning~\cite{plp,icde,sws}. Since hypergraphs represent higher-order relationships among objects
via hyperedges, Zhou et al.~\cite{zhou2006learning} introduces a new clustering objective, hypergraph normalized cut~\cite{zhou2006learning},
to solve the hypergraph clustering problem. Recently, an efficient hypergraph clustering method
called hGraclus has been proposed~\cite{whang2020mega}, and was shown to outperform state-of-the-art hypergraph clustering algorithms: our formulation of test coverage using hypergraphs is motivated by hGraclus~\cite{whang2020mega}. To the best of our knowledge, our work is the first work
that uses hypergraph modelling and clustering to solve the failure clustering problem.

\section{Conclusion and Future Work}
\label{sec:conclusion}
We design \textit{hdist}, a novel hypergraph-based distance metric for test 
cases, and propose a failure clustering technique \name, that combines 
\textit{hdist} with AHC and the distance-based stopping criterion. Our 
technique accurately clusters 68\% and 73\% of the studied Java and C 
multi-fault subjects, respectively, and outperforms other distance metrics, 
such as Jaccard, as well as the state-of-the-art failure clustering method, MSeer.
In terms of the FL performance after failure clustering,
the use of \name allows us to find the greatest number of faults, with 
less inspection cost than other methods that found a similar number of 
faults.
Our empirical evaluation shows that \name can outperform 
state-of-the-art failure clustering techniques. For future work, we will 
investigate overlapping clustering algorithms to relax the 
failure-to-single-fault assumption. Furthermore,
other higher-order relationships between test cases, such as similarity in mutation coverage
or failure history, can also be encoded by hypergraphs. We expect incorporating richer information
can produce more accurate failure clustering.

\bibliographystyle{ACM-Reference-Format}
\bibliography{ref}

\end{document}